\documentclass[12pt,preprint]{aastex}

\newcommand{\HI}{\mbox{\rm H{\sc i}}}
\newcommand{\CII}{\mbox{\rm C{\sc ii}}}
\newcommand{\CIV}{\mbox{\rm C{\sc iv}}}
\newcommand{\SiII}{\mbox{\rm Si{\sc ii}}}
\newcommand{\SiIII}{\mbox{\rm Si{\sc iii}}}
\newcommand{\AlII}{\mbox{\rm Al{\sc ii}}}
\newcommand{\MgI}{\mbox{\rm Mg{\sc i}}}
\newcommand{\MgII}{\mbox{\rm Mg{\sc ii}}}
\newcommand{\FeII}{\mbox{\rm Fe{\sc ii}}}

\begin{document}

%%%%%%%%%%%%%%%%%%%%%%%%%%%%%%%%%%%%%%%%%%%%%%%%%%%%%%%%%%%%%%%%%%%%%%%%
%%%% title page
%%%%%%%%%%%%%%%%%%%%%%%%%%%%%%%%%%%%%%%%%%%%%%%%%%%%%%%%%%%%%%%%%%%%%%%%

\title{A Uniform analysis of the Lyman alpha forest at z = 0-5:\\
III. HST FOS Spectral Atlas.}

\author{Jill Bechtold\altaffilmark{1},Adam Dobrzycki\altaffilmark{2}, \\
Brenda Wilden\altaffilmark{1,3}, Miwa Morita\altaffilmark{1},
Jennifer Scott\altaffilmark{1},
Danuta Dobrzycka\altaffilmark{2}, 
Kim-Vy Tran\altaffilmark{1,3}, 
Thomas L. Aldcroft\altaffilmark {2}}

\altaffiltext{1}{Steward Observatory, University of Arizona, Tucson,
AZ 85721, USA\\
e-mail: [jbechtold,jscott]@as.arizona.edu}

\altaffiltext{2}{Harvard-Smithsonian Center for Astrophysics,
60 Garden Street, Cambridge, MA 02138, USA, e-mail:
[adobrzycki,ddobrzycka,aldcroft]@cfa.harvard.edu}

\altaffiltext{3}{Present address: Board of Astronomy and Astrophysics,
University of California, Santa Cruz, Santa Cruz, CA 95064, USA,
e-mail: [bwilden,vy]@ucolick.org}

\slugcomment{Accepted for publication in Ap.J.Suppl.}

%%%%%%%%%%%%%%%%%%%%%%%%%%%%%%%%%%%%%%%%%%%%%%%%%%%%%%%%%%%%%%%%%%%%%%%%
%%%% abstract
%%%%%%%%%%%%%%%%%%%%%%%%%%%%%%%%%%%%%%%%%%%%%%%%%%%%%%%%%%%%%%%%%%%%%%%%

\begin{abstract}

We analyzed the absorption line spectra of all quasars
observed with the high resolution gratings of 
the {\it Faint Object Spectrograph} on board the {\it Hubble Space
Telescope}.   We examined 788 spectra for 334 quasars, and
present line lists and identifications of
absorption lines in the spectra of 271 of them.  
Analysis of the statistics of the Ly-$\alpha$ and metal 
absorption systems are presented in companion papers 
(Dobrzycki {\it et al}. 2001; Scott {\it et al}. 2001; 
Morita {\it et al.} 2001).  
The data and several analysis products 
are available electronically and on the authors' web site.
\end{abstract}

%%%%%%%%%%%%%%%%%%%%%%%%%%%%%%%%%%%%%%%%%%%%%%%%%%%%%%%%%%%%%%%%%%%%%%%%
%%% body
%%%%%%%%%%%%%%%%%%%%%%%%%%%%%%%%%%%%%%%%%%%%%%%%%%%%%%%%%%%%%%%%%%%%%%%%

\section{Introduction}

The {\it Faint Object Spectrograph} ($FOS$, Keyes et al. 1995, and 
references therein) 
on board the {\it Hubble Space Telescope}
provided the first opportunity to study quasar spectra  
in the ultraviolet with sufficient spectral resolution to 
measure the properties of the narrow absorption lines in detail.
The Quasar Absorption Line Key Project used the {\it FOS} to 
make a comprehensive
study of absorber properties at low and intermediate redshift (Bahcall et al. 
1993; Bahcall et al. 1996; Jannuzi et al. 1998, hereafter J98,
 and references therein).    
There were many quasars observed with the 
$FOS$, however, for purposes other
than the study of their absorption line properties.  We have
retrieved these data from the Space Telescope Science Institute 
archive, as well as the original Key Project data, and analyzed the
absorption lines in all of the spectra, in a uniform way.

The resulting data set is presented in this paper, and is valuable 
for many studies.  These spectra can be combined with spectra obtained
from the ground at moderate resolution (Bechtold 1994, Scott et al. 2000a and 
references therein), to provide statistics on 
the Ly-$\alpha$ forest from $z=0$ to 5 (Dobrzycki et al. 2001).  Although
line blending must be accounted for with care, these data are a useful 
complement to high spectral resolution echelle 
studies which to date have been carried
out on a much smaller number of quasars.   

In companion papers, we present 
a statistical analysis of the Ly-$\alpha$ forest lines (Dobrzycki et al. 2001);
we analyze the proximity effect in this sample, 
in order to derive a limit on the metagalactic ultraviolet background 
and its evolution (Scott et al. 2001);  and analyze the interstellar
lines from the Milky Way (Morita et al. 2001).  
Other topics will be discussed in future papers. 

In section~2 we present the data
collected from the $HST$ archive and describe the data reduction
procedure.  In section~3 we discuss individual objects. 
The spectra, continuum fits, linelists, and other analysis products
are available electronically and on the authors' web site, described in section~4.

%%%%%%%%%%%%%%%%%%%%%%%%%%%%%%%%%%%%%%%%%%%%%%%%%%%%%%%%%%%%%%%%%%%%%%%%

\section{The data}

\subsection{Reductions}

We have collected $HST/FOS$ quasar spectra taken with the G130H, G190H,
and G270H gratings. To the best of our knowledge, we have obtained 
all available quasar spectra with these gratings. Table 1
is an abbreviated list of all of the data sets we retrieved.  
A full list is given in Table 2.  Table 2 lists the 
identification numbers of the individual data sets retrieved
from the archive and their dates of observation.  Several of the spectra
turned out to be of very low quality, or were failed observations,
and were not reduced further.  Broad absorption line quasars and
objects observed in SPECTROPOLARIMETRY mode were not
reduced. These are noted in Tables 1 and 2.

Some of the data were retrieved from the archives before the best 
available calibration products had been applied to the products in 
the archive.  We recalibrated these spectra, using STSDAS.
Spectra observed prior to the installation
of COSTAR in December 1993 were reduced from 
raw, uncalibrated data, while for the
post-COSTAR observations we used calibrated images. The new
calibration methods available for pre-COSTAR data are mainly concerned
with improved scattered light corrections and flux calibration. For
all the raw data we applied the {\em addnewkeys} and {\em calfos}
tasks from the {\em stsdas.hst\_calib.fos} package, creating
calibrated files containing data similar in format and contents to the
post-COSTAR calibrated data.  The data that was re-calibrated, and the 
flatfield files used are listed in Table 2.

We then extracted the data from the reduced images following a
procedure dependent on the observation mode. If the spectrum was taken
in the RAPID--READOUT mode, we combined the spectra from all groups
within the image using the {\em rcombine} task from the {\em
stsdas.hst\_calib.ctools} package, propagating the error and data
quality flags. If the data were taken in the SPECTROSCOPY mode, the
last group in the image is the sum of all groups.

Flux and error spectra of the same object taken in succession with the
same observing setup (i.e.\ having the same detector, observation
mode, aperture and disperser) were then combined into a single
spectrum, weighting them by the exposure time of each individual
spectrum. Several objects were observed more than once with the same
grating but with different apertures. For those objects we 
analyzed the higher signal-to-noise spectrum, which typically was also
the higher spectral resolution spectrum. 
Table 2 notes which data sets were combined into the spectrum we
used for the absorption line analysis.

Spectra were dereddened using IRAF task {\em deredden}, taking as
input the \HI\ column density calculated using the COLDEN code,
written by J.~McDowell, which utilizes data from Stark et al.\ (1992);
for objects with $\delta < -40^\circ$ we used the data from the
galactic reddening map of Burstein $\&$ Heiles (1982).
Conversion from \HI\ column density to E(B-V) was assumed to 
be N$_{HI}$/E(B-V)=$4.8\times10^{21}$ cm$^{-2}$ magn$^{-1}$ (Bohlin,
Savage \& Drake 1978). 
The Milky Way reddening curve of Cardelli, Clayton \& Mathis (1989)
 was assumed.  The value of E(B-V) which was used is listed in Table 2. 

We put all the spectra and line lists on an absolute
wavelength scale where possible 
by calculating the average offset of the detected
Galactic interstellar lines in each spectrum. We looked for the
\SiII(1190,1193,1260,1526), \SiIII(1206), \CII(1334), and
\CIV(1548,1550) absorption in the G130H spectra, the \AlII(1670) line in the
G190H spectra, and \FeII(2344, 2374, 2382, 2586, 2600), \MgII(2796,2803),
and \MgI(2853) lines in the G270H spectra. In several cases,
especially for G190H data, no Galactic absorption lines were detected
and we then applied no wavelength shift to the
spectrum.  We list the offsets adopted in Table 2.

The Key Project went to great lengths to identify possible
residual flat-field features in their spectra, which would
erroneously be identified as absorption lines (see Schneider et al. 1993, 
J98).  We did not attempt to repeat this for all the spectra here. 
Therefore, there is a possibility that any particular weak absorption
line is not real.  In J98, they present 117 non-BAL quasar spectra, 
and report 88 flatfield features:  they found no flatfield features 
in 10 G130H spectra, 
47 flatfield features in 46 G190H spectra and 32 flatfield 
features in 61 G270H spectra. 
For most purposes, these features have a small effect 
statistically on the absorption line sample (see also Dobrzycki et al. 2001).

\subsection{Absorption Line Analysis}

For all spectra we determined signal-to-noise ratios, performed
continuum fits, detected absorption lines and identified the metal
systems using program FINDSL (Aldcroft, Bechtold, $\&$ Elvis 1994).
The estimations of equivalent width were done both by direct integration
of the profile, and by fitting gaussians. 
Direct integration overcomes the problem of asymmetric
lines which is seen for the large aperture in $FOS$, pre-COSTAR
(see section 2.4).
In the end, we decided {\it not} to use these spectra in the 
subsequent analysis of
the absorbers, and so used the equivalent widths from gaussian fits 
(see Paper IV).
We quantified the effect of line blending and completeness
by generating simulated data sets, and describe the results in
section 2.6. 

The significance of each absorption line was estimated in two ways.
To search for candidate absorption lines, we ran a boxcar of width 
equal to 2.46 times the full-width-half-maximum of the spectral line
spread function in pixels. We calculated the
equivalent width and its error, derived from the straightforward propagation
of errors from the  
errors on the flux in each individual pixel.  One measure of
the significance of a line is then the ratio of the equivalent width to its
error.  This underestimates the significance, however, since
the signal-to-noise in the bottom of the absorption line is worse
than it would have been had the line not been there.  We calculated
a threshold for detecting lines by calculating the one-sigma 
error for an absorption line as a function of wavelength for the
observed flux and our continuum fit in boxcars of the same width
as our search technique (9 pixels for most of the data).  
To interpolate over the
regions where there were actual absorption lines, we fit a spline
to the calculated threshold as a function of wavelength and, to 
be conservative, forced it to outline the upper envelope of the
data.  Then for each line for which the ratio of equivalent
width to its error was greater than 3 after the initial search, 
we interpolated the spline fits to the threshold files and calculated 
a significance equal to the ratio of the measured equivalent width
to the one-sigma error in the smoothed threshold file.
We kept lines where this ratio was greater than 3.5$\sigma$. The 
significance derived in this way is reported in the line lists. Any line
in these lists is real, but the sample is not complete unless
a higher threshold is applied.  We discuss completeness in section 2.6. 

\subsection{Identification of Metal Lines}

Metal absorption line systems were searched for in a number of ways.
First we reviewed the literature and examined all reported 
redshift systems, using the NASA/IPAC Extragalactic Database (NED)
 and the York et al. (1991) absorption line
catalog.  We adopted 
the search list of strong transitions given by Morton et al. (1988).  
In a few cases, we decided that redshift systems reported in
the literature were spurious;  these are noted individually in
section 3.  We also examined the lines shortward of 
Ly-$\alpha$ with equivalent widths greater than 1 \AA, and 
tested whether a plausible identification could be made if these
lines were Ly-$\alpha$. 

Finally, we used a computer program to search for redshift 
systems automatically.  We used the Morton search 
list and looked for redshifts which resulted in at least 
5 matches in the line lists.  We ran the code on simulated
Ly-$\alpha$ forest spectra with no metal systems to assess 
the probability of chance matches in the forest.  We found
that systems with 4 matches and plausible line ratios were
commonly found.  We made no assumptions about ion ratios,
but did require that all the transitions for a particular ion 
have physically reasonable equivalent width ratios given the f-values
listed in Morton (1991), and that Ly-$\alpha$ be
detected if it was in the spectral coverage.  

Since our goal was to construct a sample for studying the Ly-$\alpha$
forest, we were generous in our identifications: if there was
any doubt, we went ahead and identified a line as a metal transition.
We kept track of the regions of the spectrum
which contained metal lines, and provide this information electronically.
In effect these are parts of the redshift path which were not observed 
for Ly-$\alpha$ and our analysis programs simply delete these parts
of the spectrum automatically.  Therefore, before drawing conclusions
about any particular line identified as a metal, one should 
look carefully at all the lines identified for that redshift system.

Note that this data is {\it not} appropriate for deriving
certain types of statistics 
for the metal-line absorbers.  We didn't keep track 
of why a particular quasar was observed, and clearly some objects
were observed because they were known beforehand to have interesting
metal-line systems, for example damped Ly-$\alpha$ absorbers.  

In Figure 1, we plot the spectra which were analyzed. 
For objects which 
were observed more than once, we show the spectra
from which we derived absorption line lists.  Other spectra are available on
our web site. 

The absorption line lists, including identifications, are  
given electronically in Table 3, and on the web site.

\subsection{Effect of Aperture Size on Line Detection and Measurement}

Our dataset is a compilation of spectra taken 
in different apertures.  Post-COSTAR the spectral line spread 
function (LSF) was similar for data taken in each aperture,
and we neglected any differences.  Pre-COSTAR,  
the LSF was significantly degraded and non-Gaussian 
for data taken in the A-1 (4.3 \arcsec) aperture (Smith $\&$ Hartig 1989; 
Evans 1993). The LSF in
other apertures was similar to the post-COSTAR values. 

Seventy-two of the objects in the sample were observed pre-COSTAR with A-1,
and were not observed subsequently;  these are listed in Table 4.
In a few cases, however, observations were taken pre-COSTAR with A-1 and
post-COSTAR or with a smaller aperture, with the same 
grating, so we could directly compare the line lists
generated.  For two objects, q1521+1009 and q1634+70437, we examined
the G270H spectra in detail.  For q1521+1009, 8 lines which were 
detected in the small aperture data were strong enough to have been
expected to be detected above the 5$\sigma$ equivalent width 
threshold in the pre-COSTAR A-1 data;  however only 6 were detected.  
For q1634+7037, the situation was worse:  of the 57 strongest lines 
in the small aperture data should have been detected in the pre-COSTAR
A-1 spectrum, but only 31 were.  

Figure 2 shows the 4 spectra in question. The
resolution is low in the pre-COSTAR A-1 spectra-- 50$\%$ lower 
than in the small aperture data-- and at $z\sim$1 our line search techniques
apparently have not adequately taken into account the change in sensitivity.

While the spectra observed only in the A-1 aperture, pre-COSTAR
are useful for some types of absorption line studies, we recommend that
they {\it not} be used for studies of the Ly-$\alpha$ forest. 

\subsection{Comparison with the HST Quasar Absorption Line Key Project}

A part of our sample contains absorption lines derived from the
spectra observed for by the HST Key Project. We
derived the absorption line properties independently from the Key
Project for the same data, and in this section compare line lists. 
Although we reduced the spectral data slightly differently (weighting
individual spectra by exposure time) the spectra for a few objects
we checked are identical. 

Figure 3 shows the comparison of the lines found  
by our software and the Key Project (J98).  We chose 7 objects
for detailed comparison (q0003+1553, q0743-6719, q1116+2135, 
q1241+1737, q1424-1150, q1718+4807, q2340-0339). These were chosen
randomly from the objects which had reasonable signal
to noise, contained Ly-$\alpha$ forest, did not have 
heavily blended or broad absorption features, 
and appeared to have continua that should be fit easily.  Of a total 
of 285 lines found by our software, 14 lines, or 5$\%$ of the total,
were not found by the Key Project.
In all cases, the features in
question were weak, blended with other features, or near the edge of
the spectrum.

Figure 3 shows the comparison of the equivalent widths 
and significance level for the lines which were detected in common
by both groups.  The significance
level assigned to each line shows a systematic offset however.
We assign a significance to the line which is about 20$\%$ lower  
than that assigned by the Key Project.  As a result, our line lists
are more conservative, and contain fewer lines above a given threshold.  
This difference is probably due to the fact that we adopt a fairly
wide window (2.46 times the full-width-half-maximum of the spectral  
line spread function in pixels) in which to calculate the
detection threshold, compared to J98. 

These differences are larger than we expected.  Since both groups
have automated line finding and continuum fitting algorithms, the
differences result from the slightly different approaches to line
detection. The effects are easily quantifiable in subsequent
analysis.
 
\subsection{Completeness and Line Blending}

In order to determine how reliably we find significant lines, 
we produced a set of 25 $z=0.5$ simulated Ly-$\alpha$ forest spectra and a
set of 25 spectra at $z=1$.  The code we used is described in Dobrzycki 
\& Bechtold (1996).
We modeled the Ly-$\alpha$ forest using the neutral hydrogen 
column density (N$_{ H I}$) and b-value  distributions
observed in high resolution optical spectra of high redshift 
quasars (Hu et al. 1995).  We parameterized the number of lines
$N$ as

$N \propto N_{HI}^{-\beta} exp(-(b-<b>)^2/2 \sigma_b^2) (1+z)^{\gamma}$

\noindent
and used values of $\gamma$ = 0.50 (Bahcall et al. 1993)
and $\beta = $1.46 (Hu et al. 1995). The results of this
analysis should not be sensitive to the value of $\gamma$ since the redshift
path covered in each spectrum is small.  The lower column density limits
chosen were 5x10$^{12}$ cm$^{-2}$ for $z=0.5$ and 2x10$^{12}$  cm$^{-2}$ 
for $z=1$; and the upper
column density limit used was 3x10$^{14}$ cm$^{-2}$ in both cases. The
mean Doppler parameter and width of the Doppler parameter distribution
used were $<b> = $28 km s$^{-1}$ and $\sigma_b$ = 10 km s$^{-1}$. 
The column density limits
were chosen to give the same total absorption in the simulated spectra as
is calculated in the spectra of q0850+4400 and q2145+0643, which served as
the templates for the $z=0.5$ and $z=1$ simulations, respectively. 
We used the signal-to-noise, sampling, spectral resolution,
and observed continuum of these two representative objects to
model the Ly-$\alpha$ forest spectra. 

We find
that for the $z=0.5$ simulation, at 5$\sigma$ significance, the line lists
are 100\% complete; and for the $z=1$ simulation and 5$\sigma$ significance
the line lists are 98\% complete. In this calculation, we account for line
blending by counting all simulated lines within 2.46 resolution elements
of each recovered line as a component of that recovered line.  If blending
is neglected and matches are determined on the basis of the best
wavelength match between simulated and recovered lines, the completenesses
drop to 82\% and 61\% for $z=0.5$ and $z=1$ respectively.  For lines of
3$\sigma$ significance and above, the completenesses of the $z=0.5$ and $z=1$
simulation line lists are 94\% and 96\% when blending is accounted for and
75\% and 54\% when it is not.

The simulations also revealed another interesting point.  Of the lines
$``$recovered" by our software, a small percentage were not generated by the
simulation program.  In other words, our software found some spurious
lines.  How many of these lines occur in the sample depends upon the
signal-to-noise of the data and upon the line density of the spectra, i.e.
the redshift. The percentages of these spurious lines were 5.3\% of the
3$\sigma$ lines from the $z=0.5$ simulated spectra, and 1.3\% of the
3$\sigma$ lines from the $z=1$ simulated spectra.  The higher percentage
of spurious lines in the $z=0.5$ spectra compared to the $z=1$ spectra, which
have higher line densities, can be understood to be a result of the higher
signal-to-noise in the $z=1$ simulated spectra.  The signal-to-noise
template for the $z=1$ simulations fluctuates between 12.5 and 14 across the
spectra, while that of the $z=0.5$ simulations increases smoothly from 3 to
15 from the blue to the red. None of the lines recovered at 5$\sigma$
significance were spurious.

In summary, based on the analysis of simulated Ly-$\alpha$
forest spectra, we adopt 3.5$\sigma$ as the threshold for detection of
real absorption lines, and 5$\sigma$ as the threshold for completeness.

\section{Notes on Individual Quasars.}

We discuss the individual quasars here.  After each object, 
we specify G270($n$), G190($n$), G130($n$), to indicate that $n$
spectra taken with the G270H, G190H or G130H gratings were
reduced. The emission line redshifts are taken from Hewitt $\&$
Burbidge (1993) or NED, unless otherwise noted. 

q0002-4214, $z=2.76$.
This high redshift quasar has  been the subject of several
optical absorption line studies (Sargent et al. 1979, 
Boisse \& Bergeron 1985).  We identify
several metal lines with previously known systems, at 
$z = $2.46, 2.30, 2.16, 1.99, and 1.54.  
G270.

q0002+0507, $z=1.90$. UM18. This bright quasar has a well studied IUE spectrum,
(Bechtold et al. 1984) and two previously known absorption line
systems at $z=0.8514$ and $z=1.7444$ (Young, Sargent \& Boksenberg 1982, Steidel \& Sargent 1992).
Churchill (1997) reports a weak Mg II system at $z=0.5914$. 
G270(2),G190.

q0003+0146, $z=0.234$. 
G190.

q0003+1553, $z=0.450$.  PG. The continuum spectral energy distribution of 
this quasar is very well studied. 
J98 present the FOS spectrum.  
G270,G190,G130.

q0003+1955, $z=0.025$.  MRN 335.  The absorption line
spectrum was studied in detail by Stocke et al. (1995) and Shull, Stocke \& 
Penton (1996), and four local absorbers were found.
G270,G190,G130.

%q0005-2345, $z=1.407$,  G270H spectra were obtained, but the quasar was
%not detected and we did not process the data.

q0007+1041, $z=0.089$. IIIZw2.  A well-studied low redshift Seyfert.  
A strong line at 1322\AA  may be associated Ly-$\alpha$, 
with no corresponding metal-line absorption.
G130,G270.

q0015+1612 $z=0.553$.  The FOS spectrum is discussed in detail by
Bechtold et al. (2001).
G190.

q0017+0209, $z=0.401$. 
Discovered in the LBQS survey (Foltz et al. 1989), the FOS
spectrum is presented in Turnshek et al (1997) and J98.  
They argue convincingly that metal systems are detected at 
$z =  0.5764$, 1.3389, and 1.3426, and less convincingly for
several others. Churchill (1997) reports a Mg II system at $z=0.7289$
from optical spectra.  We identified metal lines associated with 
these four systems.
G190.

q0018+2825,  $z=0.509$. The object does not appear in NED.  The
redshift is determined from strong C IV and C III] emission lines 
at 2342 \AA and 2886 \AA respectively.  A strong absorption line at 
2344 \AA is probably associated C IV blended with interstellar
Fe II $\lambda$2344.  
G270.

q0024+2225, $z=1.108$.  The $z=1.11$ associated absorber described by
Weymann et al. (1979) is confirmed.  The FOS spectrum is referred to  
in Bahcall et al. (1993), and presented in J98.  
Other metal absorption line systems are reported by J98
at $z=0.8196$ and 0.4069.
G190, G270.

q0026+1259, $z=0.142$.  The continuum spectral energy distribution of this
PG quasar has been studied extensively.
G270, G190, G130(2).

%q0031m7042, $z=0.363$.  The spectra were noisy for this object and
%the data were not reduced.
%
%q0038+3242, $z=0.583$.
%The data were very low signal-to-noise and the data were not reduced.

q0042+1010, $z=0.583$.  Radio loud quasar with no previously published absorption line spectrum. A very strong line at 1926 \AA 
may be associated Ly-$\alpha$ 
and no obvious corresponding metal absorption.
G270, G190.

q0043+0354, $z=0.384$. The FOS spectra and the absorption line analysis
are presented earlier in Turnshek et al. (1997) and Turnshek et al.
(1994).  Turnshek et al. (1997) describe the C IV absorber at $z=0.38$ 
as a $``$BAL" although it is very narrow and would be what we call in
this paper an $``$associated absorber".  Semantics aside, there is 
a strong absorption line system at the emission line redshift 
 with Ly-$\alpha$ and C IV 
detected.
G270, G190.

q0044+0303, $z=0.624$.  
The FOS spectra are previously discussed in Bahcall et al. (1993).
They describe C IV at $z=0.245$ and $z=0.449$.  The $z=0.449$ system may
be may be identified
with an intervening  galaxy at the same redshift
(Ellingson, Green \& Yee 1991).
The evidence for the $z=0.245$ spectrum is weak, given that only the C IV
doublet is seen in the Ly-$\alpha$ forest part of the spectrum, so
the probability of a chance coincidence is high.
G270, G190.

%q0050-2523, $z=2.159$. The spectrum of this object did not have sufficient 
%signal to warrent reductions.

q0050+1225, $z=0.061$.  IZw1.  
The discovery of very local Ly-$\alpha$ absorption in the
spectrum is described by Shull, Stocke \& Penton (1996) and Stocke et al (1995).
Two systems, at $z=0.0538$ and 0.05061 appear to have Ly-$\alpha$ and
metal line systems detected.
G270(3), G190(2), G130.

q0052+2509, $z=0.155$. 
G190.

%q0053-0119, $z=0.170$. The spectrum of this object did not have sufficient
%signal to warrent reductions.

q0058+0155, $z=1.954$. PHL 938.
We carried out an analysis of this spectrum, and 
identified a number of metal-line absorbers, at $z = 0.612, 1.20, 1.26, 
and 1.46$. 
G190.

%$q0059-2735, $z=1.595$.
%$This is a BAL quasar and was not analyzed.

q0100+0205, $z=0.394$. UM 301. 
G190(2).

q0102-2713, $z=0.780$. This is an LBQS quasar (Morris et al. 1991), 
and has a strong intervening C IV absorber at $z = 0.48$.
G190.

%q0103-2622, $z=0.776$.  
%The target acquisition for this object failed, and there is no
%signal in the data. 

q0107-0235, $z=0.948$.  This is the brighter of the pair of LBQS quasars
(Chaffee et al. 1991)
whose FOS spectrum is discussed in detail by Dinshaw et al. (1995, 1997).
G190.

q0107-0234, $z=0.942$. This is the fainter of a pair of LBQS quasars (Chaffee 
et al. 1991) whose FOS 
spectrum is discussed in detail by Dinshaw et al. (1995, 1997).
G190.

q0107-1537, $z=0.861$.
A strong C IV system is at $z=0.548$.
G270.

%q0110+2942, $z=0.363$.  
%The spectrum for this object had very low signal-to-noise
%and was not reduced.

q0112-0142, $z=1.365$.  This is a well-studied Parkes radio-loud quasar, whose
FOS spectrum was presented by Wills et al. (1992).  We identify a
candidate intervening damped Ly-$\alpha$ absorber at $z=1.19$.
G270.

%q0113+3249, $z=0.016$.  MRN 1. 
%The spectra of this object do not contain the Ly-$\alpha$
%part of the spectrum and we did not analyze it.
%G270.

q0115+0242, $z=0.672$. PKS.
G270, G190.
 
q0117+2118, $z=1.493$.  This well-studied PG quasar has optical
absorption line spectra reported by Boisse et al. (1992), Steidel \& Sargent (1992)
and Rao, Turnshek \& Briggs (1995).  We identify lines from the $z=0.576$, 1.047,
1.325 and 1.342 discovered by these authors.  There is a strong
associated absorption complex at $z=1.447 - 1.506$.
G270(2), G190.

q0119-0437, $z=1.925$.
This high redshift quasar has a 
well-studied optical absorption line spectrum (Sargent, Boksenberg \& Young 1982, 
Sargent, Steidel \& Boksenberg  1988a).
We identify lines associated with several intervening metal line absorption
redshift systems.
G270, G190.
 
q0121-5903, $z=0.047$.  Fairall 9.
The Milky Way interstellar absorption is extremely  
strong, particularly the Si II 1260 line.
G270, G190, G130.

q0122-0021, $z=1.070$.  The FOS spectra were discussed by Bergeron et al. 
(1994) and Bahcall et al. (1996).  They identify intervening
metal line absorbers at $z=0.398$ and $z=0.953$.
G270, G190.

q0125-0635, $z=0.005$. 
This is the Sy 1, MRK 996. 
G190.

q0133+2042, $z=0.425$.  This is 3C47, whose FOS spectrum is described by
Wills et al. (1995).  There is a strong associated C IV absorber.
G270, G190.

q0137+0116, $z=0.260$.  UM 355.   This is a well-studied
radio-loud quasar.  
G270, G190, G130.

q0143-0135, $z=3.124$.  UM 366. This is a high redshift
quasar, detected to 581 \AA in its rest frame.  
A damped Ly-$\alpha$ absorber is detected at $z=1.61$, 
as well as lines from systems at $z=1.58$ and 1.28.  
G270.

q0150-2015, $z=2.139$.  This quasar has a well studied absorption line
spectrum (Sargent, Steidel \& Boksenberg 1988b, Steidel \& Sargent 1992,
Hamann 1997 and references therein), including a strong intrinsic absorber.   
G270, G190.

%q0151+0433, $z=0.404$.  The observation of this object had
%low signal-to-noise and was not analyzed.

q0159-1147, $z=0.669$.  3C57.  The FOS spectra is described by 
Wills et al. (1992) and J98.
A strong absorption line at the Ly-$\alpha$ emission 
line redshift (2028 \AA) may be an associated metal-line 
absorber, although no 
associated Mg II absorption is reported by Aldcroft, Bechtold
and Elvis (1994).  The part of the FOS spectrum where C IV 
is expected coincides with strong lines of interstellar Fe II,
and is inconclusive.  Likewise several candidate metal absorbers
are suggested by J98 but none are secure.
G270, G190.

%q0200-0858, $z=0.770$.  We did not reduce the spectrum. 

q0207-3953, $z=2.813$. This is a well-studied high-redshift quasar,
with a strong Lyman-limit at 3190 \AA, which is 836 \AA in its rest frame.
G270.

q0214+1050, $z=0.408$. PKS.  This is a  radio loud quasar, with no previous
absorption line spectrum published. 
G270, G190.

q0219+4248, $z=0.444$.  3C 66A.  This is a BL Lac object.  A number of
weak absorption lines are seen, but none more significant than
5$\sigma$.  No broad emission lines are seen.
G270, G190, G130.

%q0226-1024, $z=2.256$.  This is a BAL QSO and we did not analyze the
%spectrum.

q0232-0415, $z=1.450$.  This Parkes radio loud quasar has been the subject
of numerous continuum and optical absorption line studies.  The
$z=1.425$ C IV absorber reported by York et al. (1991)
is not confirmed, since a strong Ly-$\alpha$ line is not apparent.
We tentatively identify a system at $z=1.355$.  J98
discuss the FOS spectrum, and also identify several possible
metal absorbers, but none are convincing, since they rely on only 
two identified lines, in confused regions of the spectrum.
G270.

%q0237-2322, $z=2.223$.
%This is a well-studied PKS quasar.  The observation for this quasar
%failed to acquire it, and there is no signal in the spectrum.

%q0240+0044, $z=0.569$.  The signal-to-noise was very low for this
%object and we did not reduce its spectrum.

q0253-0138, $z=0.879$.  This is an LBQS quasar (Chaffee et al. 1991)
whose absorption line spectrum is not previously reported.  Only one
significant line is seen, identified with ISM Mg II $\lambda$2793 absorption.
G270.

q0254-3327b, $z=1.915$.  The quasars q0254-3327 a,b and c have been
the subject of numerous studies.  
G270.

q0254-3327c, $z=1.863$. 
G270, G190.

%q0256+3637, $z=0.012$.  MRK 1066. This spectropolarimetry observation
%was not analyzed.
%G270(2).

q0302-2223, $z=1.40$.
There is a damped Ly-$\alpha$ absorber at $z=0.99$.  
We identify lines associated with several other systems.
G270.

%q0311+4151, $z=0.024$.  MRK 1073.  The spectrum is a spectropolarimetric
%observation and we did not reduce it.
 
%q0318-1937, $z=0.104$.  The signal-to-noise was very low for this
%object and we did not reduce the spectrum.

q0333+3208, $z=1.258$.  
This is a very well studied radio-loud quasar.
We confirm the Mg II absorber at $z=0.953$ described by Steidel \& Sargent (1992).
G270.

q0334-3617, $z=1.100$.  This is an X-ray selected quasar (Maccacaro et al. 1991,
Stocke et al. 1991a).   The strong absorption line at 2557 \AA near
the Ly-$\alpha$ emission line does not appear to have any metal absorption
in C IV associated with it.
G270.

q0349-1438, $z=0.616$. 3C95.  The FOS spectrum is presented by
Bahcall et al. (1993).  One metal-line absorber is seen, with 
$z = 0.3566$.
G270, G190.

q0350-0719, $z=0.962$.  This is an interesting radio-loud quasar with  
a broad associated absorption line.  See Aldcroft, Bechtold \& Elvis
(1994) for optical spectra. 
G270.

q0355-4820, $z=1.005$. The FOS spectrum is discussed by Hamann, Zuo \& Tytler (1995).
We identify a strong metal system at $z=0.987$.
G270(2).

q0403-1316, $z=0.571$. The FOS spectrum of this well-studied Parkes
radio quasar is presented by Wills et al. (1995).  
G270, G190.

q0405-1219, $z=0.574$.  This quasar was discovered by IRAS (Low et al. 
1988) and has a strong Ly-$\alpha$-Ly-$\beta$ pair at
$z=0.6276$.  No metal transitions are seen
associated with this system however.
The FOS spectrum is discussed in detail by Bahcall et al. (1993),
with an updated discussion of the metal absorbers in J98.
One of the G130H spectra was obtained in SPECTROPOLARIMETRY mode 
and was not reduced.
G270, G190, G130(2).

q0414-0601, $z=0.781$. 3C110. 
The FOS spectrum is presented by J98.
They do not report any metal-line systems, but we find a fairly
convincing one at $z=0.664$.
G270(2), G190(2).

q0420-0127, $z=0.915$. PKS.
This is a well-studied quasar, with an optical absorption line
spectrum reported by Aldcroft, Bechtold \& Elvis (1994).
G270.

q0421+0157, $z=2.044$. PKS.
The optical absorption line spectrum
of this high redshift quasar is discussed by Steidel \& Sargent (1992).
G270.

q0424-1309, $z=2.159$. PKS 0424-13.
This optical absorption line spectrum of this high
redshift quasar is discussed by Steidel \& Sargent (1992).
G270.

q0439-4319, $z=0.593$. PKS. A low redshift damped Ly-$\alpha$
absorber at $z=0.101$ is identified
by Petijean et al. (1996), who present the FOS data.
In addition, we find another metal absorber at $z=0.44$.
The FOS spectra are also discussed by J98.
G270, G190.

q0450-2958, $z=0.286$.  This IRAS selected quasar (Low et al. 1988)
has been the subject of several studies.  A strong assoicated
C IV absorption trough is seen.
G190.
 
q0453-4220, $z=2.66$.
This is a well studied high redshift quasar, with an optically 
thick Lyman limit at 3120 \AA.
G270, G190.

q0454-2203, $z=0.534$. PKS. 
This is a PKS object whose optical absorption line
spectrum is discussed  by Aldcroft, Bechtold \& Elvis (1994).
Metal absorbers are found at $z=0.474$ and 0.482.
G270(2), G190(2), G130.

q0454+0356, $z=1.345$.  PKS. In this well-studied Parkes object
we identify lines associated with
previously known redshift systems at $z=1.15$, 1.06, 0.85.
G270(2), G190.

q0506-6113, $z=1.093$. This is a blazar with a strong associated
metal absorber at $z=1.079$.
G270.

%q0513-0012, $z=0.033$.  AKN 120.  This is a well-studied low redshift 
%Seyfert galaxy.  We did not reduce the spectrum.

q0518-4549, $z=0.0351$. PKS, Pictor A.  
This is a well-studied low-redshift radio galaxy.
G270, G190, G130.

q0537-4406, $z=0.894$. 
A well-known blazar and gamma ray source, classified as a BL Lac.  
Strong emission lines are seen. 
G270.

q0624+6907, $z=0.374$.  
An extremely luminous object found in the Hamburg survey (Reimers et al.
1995).  
The FOS spectrum is discussed by J98.
G270(2), G190(2), G130.

q0637-7513, $z=0.656$.  We identify metal absorption at $z=0.152$ and 
0.416.  The spectrum is also discussed in J98.
G270, G190.

q0710+1151, $z=0.768$.  3C 175. 
A radio-loud quasar whose optical absorption line
spectrum is discussed by Aldcroft, Bechtold \& Elvis (1994).  The FOS spectrum is
described by Wills et al. (1995).  One intervening metal absorber is
identified at $z=0.463$.
G270, G190.

q0740+3800, $z=1.063$.  3C 186. A radio loud quasar whose FOS spectrum is
presented by Wills et al. (1995).  We identify lines of intervening metal
line systems at $z=0.885$ and an associated absorber at $z=1.069$.
See also Aldcroft, Bechtold \& Elvis (1994).
G270.

q0742+3150, $z=0.462$.  PKS. Another radio loud quasar, with an associated
C IV absorber.  A low redshift Mg II system at $z=0.1917$ suggested
by Boisse et al. (1992) is not confirmed by the FOS spectrum.
The spectrum is also discussed by J98.
G270, G190.

q0743-6719, $z=1.511$. PKS.  A bright radio-loud quasar.
A number of C IV absorbers are detected.
The spectrum is discussed by J98.
The G190H and one of the G270H spectra were taken in SPECTROPOLARIMETRY
mode and were not reduced
G270(2), G190.

%q0749+2550, $z=0.446$.  No signal in the spectrum, and we did 
%not reduce it.

q0823-2220, $z=0.910$. PKS.
This is a BL Lac object, with redshift limit determined by the strong
absorption line system at $z=0.910$ (Falomo 1990).
No emission lines are seen in the FOS spectrum.
G270.

q0827+2421,  $z=0.939$. B2. 
A strong absorption line system is seen at 
$z=0.52$ (Ulrich \& Owen 1977). 
G270.

%q0830+1133, $z=2.976$.  
%No flux was detected in the spectrum of this high redshift
%quasar.

q0838+1323, $z=0.684$.  3C 207. 
There is a strong associated C IV system at $z=0.68$.
G270, G190.

q0844+3456, $z=0.064$. Ton 951.  Well studied PG quasar.  An 
associated absorber is seen.
G270, G130.

q0848+1623, $z=1.936$.
This is a high redshift quasar with an optically thick Lyman limit,
and several metal lines associated with previously known
redshift systems.
G270.

q0850+4400, $z=0.513$. US 1857.  The FOS spectra were presented by 
by Bahcall et al. (1993).
G270, G190.

q0851+2017, $z=0.306$. OJ +287.
This is a well-known BL Lac object.
No emission lines were detected.  Several interstellar
absorption lines are apparent.  It is discussed in J98.
One of the G270H and the G190H spectra were SPECTROPOLARIMETRY,
and were not reduced.
G270(2), G190.

q0859-1403, $z=1.327$. PKS FOS data were presented by Wills et al. (1995).
Aldcroft, Bechtold \& Elvis (1994) present an optical absorption line spectrum.
A Mg II system at $z=0.209$ proposed by Boisse et al. (1992) is not
confirmed, with no Fe II lines detected at the expected wavelengths
at that redshift.
G270.

q0903+1658, $z=0.411$. 3C 215.
The FOS spectrum is presented by Wills et al. (1995).  
G270, G190.

%q0903+1734, $z=2.771$. 
%This high redshift BAL quasar has heavy absorption.
%The FOS spectrum was discussed by Turnshek et al. (1996).
%We did not analyze the absorption line spectrum.
%G270.

q0906+4305, $z=0.670$. 3C 216.
The FOS spectrum is presented by Wills et al. (1995).
G270, G190.

q0907-0920, $z=0.625$.
This quasar was observed twice with the FOS.  The first observation
failed, and we so we analyzed the second one.
G270(2).

q0916+5118, $z=0.553$. We identify metal absorption lines
at $z=0.51$ and 0.24.  The spectrum is discussed by
J98.  The $z=0.24$ system is not identified by these authors.
G270, G190.

q0923+3915, $z=0.699$. B2.  This is a radio loud quasar whose
FOS spectrum is discussed by Wills et al. (1995).  A strong
Ly-$\alpha$ associated absorption line is seen,  
but there is no corresponding N V.
Only one of the G190H spectra were analyzed, since the other
spectra had no signal.
G270(2), G190(2).

%q0932+5006, $z=1.920$.  This is a high redshift BAL quasar
%and we did not analyze the spectrum.
%G270.

q0933+7315, $z=2.528$. 
This high redshift quasar has a strong Lyman limit 
system at 3020 \AA, and a well studied optical
absorption line spectrum.  Most of the lines
in the FOS spectrum are associated with previously
known metal absorption line redshift systems.
G270.

q0935+4141, $z=1.97$. PG.
This bright PG quasar has been the subject of numerous studies.
A damped Ly-$\alpha$ absorption line system is seen at
$z=1.396$.  There is also a Lyman limit absorber at $z=1.4649$.
G270(2).

%q0940+5425, $z=0.006$.  The observation of this object
%had very low signal and we did not reduce it.

q0945+4053, $z=1.252$. 4C+40.24.
G270.
 
%q0946+3009, $z=1.216$. PG
%This is a BAL quasar and we did not analyze the spectrum.
%G270, G190.

q0947+3940, $z=0.206$. PG.
A bright PG quasar with mostly ISM lines.
G270, G190, G130.

q0953+4129, $z=0.239$. PG.  The FOS spectrum is
discussed by J98.  There is a strong
Ly-$\alpha$ associated absorption line with
no corresponding C IV absorption detected.
G270, G190, G130.

q0953+5454, $z=2.584$.
This high redshift quasar has an optically thick Lyman
limit at 3230 \AA.  
G270.

q0954+5537, $z=0.909$. 4C+55.17.
The FOS spectrum is presented by Wills et al. (1995).
G270, G190.

q0955+3238, $z=0.533$. 3C 232.
The optical absorption line spectrum of this object is
well studied because of its chance projected proximity 
on the sky to a nearby galaxy (Stocke et al. 1991b and 
references therein.)  Several lines associated with the
foreground galaxy are detected, as well as an associated
CIV absorber.  The FOS spectrum is discussed by J98.
G270, G190, G130.

q0957+5608A, $z=1.414$. 
This is the well-known lensed quasar, with FOS spectra
of both A and B images.  There is a strong damped Ly-$\alpha$
system.  
The FOS spectra of both objects are discussed by Michalitsianos 
et al. (1997).
G270.

q0957+5608B, $z=1.414$. 
See notes for q0957+5608A.
G270, G190.

q0958+2901, $z=0.1848$. 3C 234. This object has narrow emission
lines and what is probably an associated CIV absorber, although
confirmation would require a spectrum which extends farther
into the UV.
G270, G190.

q0958+5509, $z=1.75$. MRK 132.
A  well-studied high-redshift quasar with numerous lines from previously
known metal absorption line systems detected.
An optically thick Lyman limit cuts off the spectrum at
2500 \AA.  The spectrum presented here is somewhat
higher signal-to-noise than the one presented in J98 for 
reasons which are not clear.
G270.

q0959+6827, $z=0.773$.  
G270, G190.

q1001+0527, $z=0.161$.  PG. 
The well studied quasar has a C IV absorption trough,
at 1762 A, near the emission line redshift.
G270(2), G190(2), G130(2).

q1001+2239, $z=0.974$.  PKS.
A radio loud quasar, with no previously published absorption line
spectra.  There are a few strong lines near the Ly-$\alpha$ emission line 
which don't appear to have metal absorption associated with them.
G270.

q1001+291, $z=0.329$. Ton 28.
The FOS spectrum is presented by Bahcall et al. (1993).
A broad absorption feature is seen at C III] $\lambda$1909 in the
quasar rest frame.
G270, G190, G130.

q1007+4147, $z=0.611$. 4C 41.21
The optical absorption line spectrum is reported by Aldcroft, Bechtold \& Elvis 
(1994).
There is a metal absorption system at $z=0.389$. 
G270, G190.

q1008+1319, $z=1.287$. PG
The optical absorption line spectrum is reported by Boisse et al. (1992),
who detect Mg II systems at $z=1.159$ and 1.254. 
We detect a system at $z=0.90$.  
G270.

q1010+3606, $z=0.070$. CSO 251, Ton 1187.
A low redshift Sy 1 galaxy.
Strong absorption lines are seen at $z\sim$0.
G130.

q1017+2759, $z=1.928$. Ton 34.
A high redshift quasar, with known absorbers at several
redshifts.  The FOS spectrum is discussed in J98.
G270.

q1026-0045a, $z=1.437$.
This is a bright LBQS quasar (Hewett, Foltz \& Chaffee 1995).
There is a strong metal absorber at $z=1.2959$.
There is a strong absorption line near emission Ly-$\alpha$ which does not
appear to have any metal lines associated with it.
G270.

q1026-0045b, $z=1.53$. 
This is a bright LBQS quasar (Hewett, Foltz \& Chaffee 1995).
G270.

q1028+3118, $z=0.1782$. B2
We identify a possible low redshift Mg II absorber at $z=0.0570$.
G270, G190(2).

q1038+0625, $z=1.270$ 4C 06.41.
We identify a metal absorber at $z=1.2449$.
There is also a previously known system at $z=0.441$.
There are a number of other possible metal absorber candidates
described by J98, but they are not likely real.
G270.

q1047+5503, $z=2.165$. 
We identify several candidate metal absorbers,
but most of the spectrum is blueward of Ly-$\beta$ in 
the quasar rest frame. 
G270.

q1049-0035, $z=0.357$. PG
There are strong associated absorbers at 
$z=0.3413$ and $z=0.34877$.
The $z=0.341$ system is discussed by J98.
G270, G190.

q1055+2007, $z=1.110$. PKS.
There is a Lyman limit system reported for this object
and we see Ly-$\alpha$ at $z=1.0373$ (J98).
G270.

q1100-2629, $z=2.145$. 
This high redshift quasar has numerous absorbers identified in
optical spectra, but since the Ly-$\beta$ in the rest frame is
at 3226 \AA, we did not attempt identifications.
G270.

q1100+7715, $z=0.311$. 3C249.1.
This has an associated absorber at $z=0.308$, seen in Ly-$\alpha$
and O VI, although J98 do not identify it as such.
G270, G190.

%q1101+3828, $z=0.031$. MRK 421.
%This is a nearby BL Lac object.  No emission lines are seen.
%The only significant absorption lines are Mg II at $z\sim$0.
%We did not analyze the spectrum.
%G270, G190.

q1103-0036, $z=0.425$. PKS. This is a radio source and a PG quasar.
The FOS spectra are presented by Wills et al. (1995).
G270, G190.

q1103+6416, $z=2.19$.
This bright high redshift quasar was discovered by Reimers et al. (1995).
There is an optically thick Lyman limit at 2648 \AA, probably 
the strong metal absorber at $z=1.891$.
G270, G190.

q1104-1805a, $z=2.303$.  This is
a lensed quasar (Smette et al. 1995), and there are several prominent metal
absorbers but we did not attempt identifications since Ly-$\beta$ in
the quasar rest frame is redward of the FOS spectra.
G270.

q1104-1805b, $z=2.303$. 
See notes for q1104-1805a.
G270.

q1104+1644, $z=0.634$. 
We identify metal absorbers at $z=0.454$ and a tentative system
at $z=0.387$.  
G270, G190.

q1111+4053, $z=0.734$. 3C 254. 
There is a strong associated absorber detected in O VI and N V.
The FOS spectrum is shown by Wills et al. (1995).
G270, G190.

q1114+4429, $z=0.144$. PG. 
A strong associated absorber is seen.
G270(2), G190(2), G130(2).

q1115+0802A1, $z=1.722$. PG
Lensed quasar.  In addition to an absorber with $z=1.69$ near
the quasar redshift, we find candidate metal 
absorbers with $z=1.04$ and 0.99.
G270(2).

q1115+0802A2, $z=1.722$. PG.
See discussion of q1115+0802a1.  We identify candidate metal absorbers
with $z=1.04$ and 0.99 in this image as well.
G270(2).

q1115+4042, $z=0.154$. PG.
We identify a C IV system at $z=0.155$.  Most of the detected
lines are from the ISM.
G270, G190 G130.

q1116+2135, $z=0.117$. PG. J98.
G270(2), G190(2), G130.

q1118+1252, $z=0.685$. 
This object has a strong associated metal-line absorber.
G270, G190.

%q1120+0154, $z=1.465$. UM 425.
%This is a BAL QSO and we did not analyze the spectrum.
%G270.

q1121+4218, $z=0.234$. 
G190.

q1122-1648, $z=2.40$. HE. We identify a $z=0.68$ metal
absorption line system, described by Reimers et al. (1995).
G270, G190.

%q1123+3935, $z=1.470$. B3.  There was no signal in the data
%and we did not carry out any analysis.

q1124+2711, $z=0.378$. This quasar is close in projection to the galaxy
cluster Abell 1267, at v=9623 km/sec.  The C III] $\lambda$1909 
profile is peculiar, but the features cannot be readily identified
with absorption from the foreground Abell 1267 or other
galaxies at $z=0.0529$ which are also nearby in the sky.
G270.

q1127-1432, $z=1.187$. PKS.    The G190H spectrum had no signal
and we did not process it.  In the G270H spectrum, 
we identify lines with a $z=0.313$ 
reported by Bergeron \& Boisse (1991).  A system with $z=0.382$ also
described by them would have no lines expected in the
wavelength range covered.  There is a strong associated Ly-$\alpha$ line
but no detected metal absorption corresponding to it.
G270.

%q1129-0229, $z=0.333$.  There was no signal and we did not process the
%data.

q1130+1108, $z=0.510$. J98.  There is a strong associated 
absorber.
G270, G190.

q1132-0302, $z=0.237$.
G190.

q1136-1334, $z=0.557$.  J98. We identify a C IV system at $z=0.332$,
which is not identified by J98.  We do not identify a system they
suggest, at $z=0.4064$.
G270, G190.

q1137+6604, $z=0.652$. 3C263.0. There is previously 
reported associated absorption
for this object by Aldcroft, Bechtold \& Elvis 
(1994).  A low redshift metal absorber
is seen at $z=0.116$.  The FOS spectrum is discussed by Bahcall et al. (1993).
G270(2), G190(2).

q1138+0204, $z=0.383$. A strong associated absorber is seen.
G190.

q1144-0115, $z=0.382$. A noisy spectrum, with a C IV absorber seen 
at $z=0.37$.  
G190.

%q1144-3755, $z=1.048$. This spectrum is fairly noisy and we did not
%analyze it.

q1146+1106c, $z=1.01$.  See discussion of q1146+103b.
A probable metal absorption line system is seen at $z=0.85982$,
as well as associated O VI absorption at $z=1.01$.  The continuum
shows a broad feature in the UV of uncertain origin.
G190.

q1146+1103e, $z=1.10$.  See notes for q1146+1103b.  There is a possible
metal absorber at $z=0.79466$. 
G190.

q1146+1104b, $z=1.01$.  This is a member of a group of quasars,
within a few arcminutes of each other on the sky.
While it has been suggested in the literature that these objects
are lensed images of the same quasar, it's clear from the FOS
spectra that this is not the case.
G190.

q1146+1105d, $z=2.12$.  There was no flux in this spectrum and we did
not analyze it.  It is part of the group described in the notes
for q1146+1103b.

%q1146+1111, $z=0.863$.  This is a BAL quasar and we did not analyze
%the spectrum.
%G270, G190(2).

q1148+5454, $z=0.978$.  
G270(2), G190(2), G130.

q1150+4947, $z=0.334$.
G270, G190.

q1156+2123, $z=0.349$.  
G270, G190.

q1156+2931, $z=0.729$. 4C29.45.  There is a strong
associated absorber at $z=0.729$.
G270, G190.

q1156+6311, $z=0.594$.  This spectrum does not contain Ly-$\alpha$
and we did not analyze the absorption lines.
G270.

q1202+2810, $z=0.165$.  GQ Comae.  This is the well-known variable
quasar.  The FOS spectra are presented by J98.
G270(2), G190(2), G130.

q1206+4557, $z=1.158$. PG. J98 presented the FOS spectra. 
Metal absorption is seen at $z=0.927$.
G270, G190.

q1211+1419, $z=0.085$. PG.  Well-studied PG quasar.  
G270(2), G190(2), G130.

%q1213-0017, $z=2.69$. UM 485. This is a BAL quasar and we did not
%analyze the absorption spectrum.
%G270.

%q1214-2745, $z=0.026$. There is no signal and we did not reduce the
%data.

q1214+1804, $z=0.375$. Probable associated absorption at $z=0.3684$.
Another strong Ly-$\alpha$ line at slightly
higher redshift has no corresponding C IV absorption.
G190.

q1215+6423, $z=1.288$. 4C 64.15.
There is a metal absorber at $z=0.99544$. A complex of absorption
at the Ly-$\alpha$ emission line could be associated Ly-$\alpha$
with no obvious metal absorption or a complex of C IV absorbers
at $z=0.805$ and $z=0.810$.
G270.

%q1216-0102, $z=0.415$. This observation had low signal-to-noise
%and we did not analyze the spectrum.

q1216+0655, $z=0.334$. J98 suggest a metal line
absorber at $z=0.1247$ but there are not strong arguments supporting
it.
G270, G190, G130.

q1216+5032a, $z=1.450$.  This object and q1216+5032b are a pair
of quasars separated by 9 $\arcsec$ (Hagen et al. 1996).  
We identify a metal absorber at $z=0.98$.  
G270.

%q1216+5032b, $z=1.451$.  See note for q1216+5032a.  This object is
%a BAL QSO and we did not analyze the spectrum.
%G270.

q1219+0447, $z=0.094$.  There are metal absorption systems
at $z=0.005$ and 0.092. 
On the sky, the quasar is close to the foreground galaxy,
M61.  The FOS spectrum and the relation of the absorbers to
M61 are discussed by Bowen, Blades \& Pettini (1996).
G270, G130.

q1219+7535, $z=0.070$. MRK 205.
The FOS spectra are presented by Bahcall et al. (1992) and J98, who 
discuss Mg II absorption associated with the foreground galaxy
NGC 4319 at $z=0.0047$.  
G270, G190, G130.

q1220+1601, $z=0.081$. 
This AGN was discovered by Margon, Downes \& Channan  (1985).
G270.

q1222+2251, $z=2.046$.  PG, TON 1530.
The FOS spectra of this
 well-studied quasar are previously discussed by Impey et al. (1996).
We identified lines associated with absorption line systems at 
$z=0.6681$, 1.4867, 1.5239, 1.9372, 1.9805, and 2.0555.
The G270H and one of the G190H spectra are
SPECTROPOLARIMETRY and were not reduced.
G270, G190(2).

q1224-1116, $z=1.979$. 
No signal was detected and so we did not process it.

q1225+3145, $z=2.219$.  B2.
The FOS data for this  well-studied quasar are 
described by Stengler-Larrea et al. 
(1995), and a ground-based echelle spectrum is given by 
Khare et al. (1997).  
We identify Ly-$\alpha$ and other transitions in systems
at $z=1.227$ and 1.429 suggested by Khare et al. (1997),
but do not confirm their $z=0.365$ system.
We identify lines associated with other well-established systems
at $z=1.6257$, 1.7953, 1.8871,  1.8974.
G270.

q1226p0219, $z=0.158$. 3C 273.  
The FOS data are discussed in Bahcall et al. (1991).
3C273 was observed in SPECTROPOLARIMETRY mode also,
but we did not analyze those data.
G270, G130.

q1229-0207, $z=$1.045.  PKS.
We identify lines associated with a damped Ly-$\alpha$ absorber
at $z=0.395$ and two strong C IV systems at $z=0.756$ and $z=0.698$.
The FOS spectrum is discussed by Stocke et al. (1998).
G270, G190.

%q1229+6430,  $z=0.170$.  
%This object had no signal, and we did not process it.
 
q1230+0947, $z=0.420$.  LBQS. 
G190.

q1241+1737, $z=1.273$. PG. 
J98 lists tentative metal absorbers at 
$z=1.272$, 1.2154, and 0.9927.
We identify lines of the system at $z=1.2154$ which is
probably an associated absorber:  strong O VI and Ly-$\alpha$ are
present.
G270.

%q1244+3225, $z=0.949$.  B2.
%No signal detected, and we did not process it.

%q1245+3431, $z=2.068$.  
%No signal detected, and we did not process it.

%q1246-0542, $z=2.236$.  
%This is a BAL QSO, and we did not analyze the
%spectra.
%G270, G190.

q1247+2647, $z=2.043$. PG.
Well-studied quasar with a damped Ly-$\alpha$ absorber at
$z=1.2276$, first suggested by Lanzetta, Wolfe \& Turnshek (1995) from IUE data.
Optical absorption line spectra are presented by 
Sargent, Steidel \& Boksenberg  (1988b) and Steidel \& Sargent (1992).
We identify lines associated with absorbers at $z=1.22$,1.96 and 1.41.
G270(2).

q1248+3032, $z=1.061$.  B2.
Well-studied radio-loud quasar with no previous absorption line
spectra published.  Several Ly-$\alpha$ lines are seen with
no associated metal absorption.
G270.

q1248+3142, $z=1.02$. CSO173
Identified by Sanduleak \& Pesch (1984), this object has no
previous absorption line spectroscopy published.
We find a fairly secure metal absorber at $z=0.6642$, and
two other tentative ones at $z=0.7323$ and 0.8961.
G190.

q1248+4007, $z=1.030$.  PG. FOS data are presented by J98.
We identify lines with redshift systems at $z=0.7760$ and 0.8553.
G270, G190.

q1249+2929, $z=0.82$.  CSO176
Identified by Sanduleak \& Pesch (1984), we identify absorbers with
$z=0.4103$, 0.5959, 0.70675.
G190.

q1250+3122, $z=0.78$. CSO179
Identified by Sanduleak \& Pesch (1984), we identify absorbers with
$z=0.5877$, 0.6393, 0.6839 and 0.72686.
G190.

q1250+5650,  $z=0.321$.  3C277.1.
The FOS spectrum is presented by Wills et al. (1995).
G270, G190.

q1252+1157, $z=0.871$.  PKS. 
J98 list a metal absorber at $z=0.6395$, which we adopt.
G270, G190.

q1253-0531, $z=0.538$.  3C279.
One of the G190H spectra had no signal and was not reduced. 
G270, G190(2), G130. 

%q1254+0443, $z=1.024$.  PG.
%A BAL QSO which we did not analyze.
%G270, G190.

%q1254+5708, $z=0.042$. MRK 231.
%This well-known object has a strong BAL-like absorber 
%at the emission line redshift.  We did not analyze
%the spectrum.
%G270, G190.

%q1257+2840,  $z=0.092$. 
%The spectrum had low signal and we did not reduce it.

q1257+3439,  $z=1.375$.  B2. 
J98 report systems at $z=1.3799$ and 1.0512, and we concur.
G270.

q1258+2835, $z=1.355$. 
A strong, broad associated Ly-$\alpha$ absorber is seen.
G270.

q1259+5918, $z=0.472$.  PG. 
Bahcall et al. (1993) present the FOS spectrum.  We identify lines
with a system at $z=0.2196$.
G270, G190, G130.

q1302-1017,  $z=0.286$.  PKS. J98 present the FOS spectrum.
G270, G190, G130.

q1305+0658, $z=0.602$.  3C281.  J98 present the FOS spectrum.
G270, G190.

q1306+3021, $z=0.806$.  J98 present the FOS spectrum. 
G270.

q1307+0835, $z=0.155$.  
G190.

q1307+4617, $z=2.129$.  J98 present the FOS spectrum. 
We found metal absorption lines from systems at $z=2.08$, 1.434, and 1.306.
G270, G190.

%q1309-0536, $z=2.188$.
%The spectrum had low signal-to-noise.
%G270.

q1309+3531, $z=0.184$.  PG.
A very strong damped Ly-$\alpha$  is seen 
near the emission redshift, at $z=0.17925$.
G270, G190, G130.

q1311+0217, $z=0.306$.  
G190.

q1317-0142, $z=0.225$.
G190.

q1317+2743, $z=1.022$.  TON153. J98 present the FOS spectrum.
G270, G190.

q1317+5203, $z=1.055$.  
A broad associated absorber at $z=0.92895$ is seen.
G270.

q1318+2903,  $z=0.549$.  TON156.
G270, G190.

q1318+2903a,  $z=1.703$.  TON155.
We find  metal absorption at $z=1.139$.
G190.

q1320+2925,  $z=0.960$.  TON157.
G270, G190.

q1321+0552,  $z=0.2051$. 
G190.

q1322+6557, $z=0.168$.  PG.
G270, G190, G130.

q1323+6530, $z=1.618$.   4C65.15.
There is a damped Ly-$\alpha$ absorber at $z=1.6101$.
G270.

q1327-2040, $z=1.169$.  PKS.
A possible associated absorber is detected at Ly-$\alpha$.
G270.

q1328+3045, $z=0.849$.   3C286.0
There is a damped Ly-$\alpha$ at $z=0.690$.
G270, G190.

q1329+4117, $z=1.93$.   PG.
Damped Ly-$\alpha$ at $z=0.5193$, and possibly $z=1.282$ are seen.
G270, G190.

q1333+1740, $z=0.554$.  PG.
J98 observed this quasar.
G270, G190.

q1334-0033, $z=2.783$.
G270, G190.

q1334+2438, $z=0.1076$.
G270, G190.

q1338+4138, $z=1.219$.
Observed by J98.
One of the G190H spectra was SPECTROPOLARIMETRY and was not
reduced.
G270, G190(2).

q1340-0038, $z=0.326$.
We find a metal line system at $z=0.2267$.
G190.

q1340+6036, $z=0.961$.  3C288-1
There is a strong associated absorber.
G270.

q1346+2637, $z=0.598$. 
G270.

q1351+3153, $z=1.326$.  B2.
There is a damped Ly-$\alpha$ at $z=1.1486$, and a somewhat doubtful
system at $z=1.1486$.
G270.

q1351+6400, $z=0.088$.  PG.
A strong associated absorber is at $z=0.0820$.
G270, G190, G130.

q1352+0106,  $z=1.117$. PG.
G270,G190(2).

q1352+1819, $z=0.152$.   PG.
G270, G190, G130.

q1354+1933, $z=0.719$.  PKS.
J98 observed this object, and list a metal absorption line system
at $z= 0.4563$.  We do not confirm a previously reported 
system at $z=0.4306$.
G270(2), G190(2).

%q1354+2552,  $z=2.032$.   PKS.
%This is a BAL QSO, despite being a PKS object, 
%and we did not analyze the spectrum.
%G270.

q1355-4138,  $z=0.313$.  
G190.

q1356+5806,  $z=1.375$.   4C58-29. 
G270.

q1401+0952, $z=0.441$.
G190.

q1402+2609, $z=0.164$. PG.
G270, G190(3), G130.

%q1402+4341, $z=0.320$. 
%Turnshek et al. (1997) present the FOS spectrum.  There is an
%associated absorber.
%G270.

q1404+2238, $z=0.098$.  PG.
There is an associated absorber at $z=0.0915$.
G270, G190, G130.

q1407+2632, $z=0.944$.   PG.
J98 presented the FOS data and suggested several redshift systems.
We concur with the ones at $z=0.6828$, 0.5998, and 0.9566.
G270,G190.

%q1408+5642, $z=2.562$.
%No signal, so we did not process it. 

%q1411+4414, $z=0.089$.  PG.
%At $z=0.084$ we find an associated absorber.
%G270, G190, G130.

%q1413+1143, $z=2.551$.
%We did not analyze the absorption lines in this high redshift
%quasar.
%G270.  

q1415+4343, $z=0.002$.  SBS.
There is an associated absorber.
G190.

q1415+4509, $z=0.114$.  PG.
G270(2), G190, G130.

q1416-1256, $z=0.129$.  PG.
G190.

q1416+0642, $z=1.436$.  3C298.
This radio loud quasar has an associated absorber at $z=1.437$.
G270.

q1424-1150,  $z=0.806$.  PKS.
J98 report a metal absorber at $z=0.6553$. A previously 
suggested system at $z=0.7465$ is not confirmed.
G270, G190.

q1425+2003,  $z=0.111$.
G190.

q1425+2645,  $z=0.366$.  B2.
There is associated absorption at $z=0.3605$ and 0.3643.
G270, G190.

q1427+4800,  $z=0.221$.  PG.
There appears to be a damped Ly-$\alpha$ absorber at $z=0.1206$.
G270, G190, G130.

q1435-0134,  $z=1.310$.
G270.

q1435+6349, $z=2.068$.  S4.
Discussed by J98.  Of the several metal line systems reported in 
the literature, we find evidence for the ones at 
$z=1.4792$, 1.5925, 1.9242 and 1.0682. 
G270.

%q1439+5343, $z=0.038$.  MRK477.
%The signal was low, and we did not reduce the data.

q1440+3539, $z=0.077$. MRK478
We see an associated absorber.
G270, G190(2), G130.

q1444+4047,  $z=0.267$.  PG.
G270, G190(2), G130.

q1451-3735,  $z=0.314$.  PKS.
G190, G130.

q1503+6907,  $z=0.318$.  B2.
There is an associated absorber. 
G270.

q1512+3701,  $z=0.371$.  B2.
A previously reported $z=0.3574$ system is not confirmed.
G270, G190.

q1517+2356, $z=1.903$.   LB9612.
G270, G190.

q1517+2357,  $z=1.834$.  LB9605. 
One of the G190H spectra had no signal and was not reduced.
G270, G190(2).

q1521+1009,  $z=1.324$.  PG.
Some of the spectra were SPECTROPOLARIMETRY and were not reduced.
G270(3), G190(2).

%q1524+5147,  $z=2.860$.
%We did not reduce the data for this BAL QSO.
%G270(3).

%q1535+5443,  $z=0.039$.   MRK486.
%Possible associated absorber.
%G270, G190.

q1538+4745,  $z=0.770$.  PG.
J98 present the spectrum and suggest several 
metal line systems.
We concur with the ones at $z=0.729$, 0.7705, 0.514, 0.706, and
 0.408.
G270, G190.

%q1540+1805, $z=1.662$.  4C18.43.
%Too noisy, we didn't reduce the spectrum.

q1542+5408, $z=2.371$. SBS.
We find metal absorption at $z=1.41$, 0.1558 and 0.72
G270, G190.

q1544+4855,  $z=0.400$.
There is a previously reported system at $z=0.0749$,
and other systems we identify at $z=0.222$ and 0.187. There is a 
fairly uncertain system at $z=0.276$.
One of the G130H spectra had no signal and was not reduced.
G190, G130(2).

q1545+2101,  $z=0.264$.  3C323-1.
There is an associated absorber.
G270, G190, G130.

q1555+3313, $z=0.942$.   B2.
G270.

q1611+3420,  $z=1.401$.
A strong associated absorption with metals is seen.
G270.

q1612+2611,  $z=0.131$.
G190.

q1615+3229,  $z=1.681$.  3C332.
G270, G190, G130.

q1618+1743,  $z=0.555$.   3C334.0.
G270(2), G190(2).

q1622+2352, $z=0.927$.  3C336.0.
Damped Ly-$\alpha$ is present at $z=0.6543$ and 0.8908.
We find metal absorption at $z=0.4718$ as well.
G270, G190.

q1623+2653,  $z=2.526$.  
There is a damped Ly-$\alpha$ absorber at $z=1.0397$.
G270.

q1626+5529, $z=0.133$.   PG.
There is a strong associated absorber.
G270, G190, G130.

q1630+3744, $z=1.478$. 
G270, G190.

q1631+3930, $z=1.023$.
There is an associated absorber at $z=1.021$.
G270.

q1634+7037, $z=1.337$.  PG.
Previously discussed by J98.
The G190H and one of the G270H spectra were SPECTROPOLARIMETRY and
were not reduced.
G270(2), G190.

q1637+5726, $z=0.745$.
G270, G190.

q1641+3954, $z=0.595$.   3C345.
Some of the data were SPECTROPOLARIMETRY and were not reduced.
G270(2), G190(2).

%q1652+3950,  $z=0.033$.   MRK501.
%G270, G190.

q1656+0519,  $z=0.887$.   PKS.
G270.

%q1700+5153,  $z=0.288$.   PG.
%A BAL QSO; we did not analyze it.
%G270, G190, G130.

q1700+6416,  $z=2.722$.
The FOS spectrum is discussed by Vogel \& Reimers (1995).
The spectral coverage is of the extreme UV and so we did not
make a line list for this object.
G270, G190(2), G130.

%q1701+6102,  $z=0.164$.
%No signal, not reduced.

q1704+6048,  $z=0.371$.  3C351.0.
There is an associated absorber at $z=0.3646$.
There are many systems reported in the literature, but
we find evidence for $z= 0.2216$, 0.3172, and 0.3646 only.  
G270, G190, G130.

q1715+5331,  $z=1.940$.   PG.
J98 list $z=1.6333$. 
We do not confirm a previously reported system at $z=0.3672$.
G270.

q1717+4901,  $z=0.025$.   ARP102B.
The signal to noise was low in the G130H spectrum.
No metal identifications were attempted.
G270, G190, G130.

q1718+4807,  $z=1.084$.   PG.
J98 lists systems at $z= 0.8929$, 1.0323, and 1.0872. 
We find absorbers at $z=1.0323$, 1.0872, and 0.7012. 
G270(3), G190(2).

q1803+7827,  $z=0.680$.   S5.
G270, G190.

q1821+1042,  $z=1.360$.   PKS.
G190.

q1821+6419,  $z=0.297$. 
Associated absorber.
G270, G190, G130.

q1845+7943, $z=0.0561$.  3C390.3.
Associated absorber at  $z=0.0506$.
G270, G190, G130.

q1928+7351, $z=0.302$.   4C73.18.
G270(2), G190(2).

%q1935-6914, $z=3.152$.
%No signal, so we did not reduce the spectrum.

q2041-1054, $z=0.035$.   MRK509.
There is a strong associated absorber.
G270, G190, G130.

q2112+0555, $z=0.398$.   PG.
G270, G190.

q2128-1220, $z=0.501$.   PKS.
There is a damped Ly-$\alpha$ at $z=0.429$, see J98. 
G270, G190.

q2135-1446, $z=0.200$.   PKS.
There is an associated absorber at $z=0.200$.
G190, G130.

q2141+1730, $z=0.213$. 
Associated absorber at $z=0.2108$.
G270, G190(2), G130(2).

q2145+0643, $z=0.999$.   PKS.
There are several metal absorption systems listed in NED,
but only $z= 0.7905$, 0.7897,0.6557
and 0.8797 are confirmed.
G270, G190.

q2155-3027,  $z=0.116$.   PKS.
Although a BL Lac, the redshift is well-determined from
spectroscopy of the nebulosity of the host galaxy and host
cluster of galxies (Falomo, Pesce \& Treves 1993).
All of the spectra are SPECTROPOLARIMETRY, except one of the
G130H spectra.
G270, G190, G130(2).

%q2200-4202,   $z=0.069$.   BL Lac.
%We did not process the data.

q2201+3131,  $z=0.297$.   B2.
A previously reported system at $z=0.282$ is not confirmed.
G270, G190, G130.

q2212-2959, $z=2.703$.   PKS.
G270.

%q2215-0347,  $z=0.241$.
%Low signal, we did not reduce the spectrum.

q2216-0350, $z=0.901$.   PKS.
G270(3), G190(4).

q2223-0512, $z=1.404$.  3C446.
A previously reported system at $z=0.4925$ is not confirmed.
G270.

q2230+1128,  $z=1.037$.   CTA102.
G270.

q2243-1222,  $z=0.630$.   PKS.
G270, G190.

q2251-1750,  $z=0.068$. 
We see an associated absorber at $z=0.06329$.
G270, G190, G130.

q2251+1120, $z=0.323$.   PKS.
There is an associated absorber at $z=0.3256$.
A previously reported system at $z= 0.2633$ is not confirmed.
Some of the spectra are SPECTROPOLARIMETRY and were not reduced.
G270(2), G190(2), G130(2).

q2251+1552,  $z=0.859$.  3C454.3.
Of several metal systems previously reported, we confirm
those at $z=0.1538$, 0.3906, and 0.8137.
G270(3), G190(3).

q2300-6823, $z=0.512$.   PKS.
G270, G190.

%q2302+0255, $z=1.052$.   PG.
%Interesting BAL-like feature at $z=0.695$-- see Jannuzi et al. (1996). 
%G270(2), G190.

q2308+0951, $z=0.432$.  PG.
A weak associated absorber at $z=0.434$ is seen.
G270, G190, G130.

q2340-0339,  $z=0.896$.   PKS.
J98 list a number of candidate metal redshifts.
We use those at $z=0.6841$, 0.4621, 0.4212, and 0.8238.
One of the G270H spectra was not reduced because it had no signal. 
G270(3), G190.

q2344+0914, $z=0.672$.   PKS.
J98 lists $z= 0.1176$ and 0.4368 for metal systems; we used $z=0.4368$
only.
G270, G190.

q2347-4342, $z=2.90$.
G270, G190.

q2349-0125, $z=0.174$.
G190.

q2352-3414, $z=0.706$.   PKS.
Discussed in J98.
G270, G190.

%%%%%%%%%%%%%%%%%%%%%%%%%%%%%%%%%%%%%%%%%%%%%%%%%%%%%%%%%%%%%%%%%%%%%%%%

\section{Description of Data Products and Web Site.}

In addition to the electronic version of the tables and figures
published here, we have put other electronic
files on web sites at the Steward Observatory, at the 
University of Arizona, and the High Energy group at the Center for
Astrophysics, see 

http://lithops.as.arizona.edu/$\sim$jill/QuasarSpectra or 

http://hea-www.harvard.edu/QEDT/QuasarSpectra.  

We put there the
reduced spectra shown in Figure 1, as well as other spectra which
we reduced but did not use to construct absorption line lists.
Data are available as ascii lists, or fits files.
We also put the line lists with identifications of metal absorption lines, 
and the equivalent width threshold for detection 
as a function of wavelength for each analyzed spectrum.   
A detailed description of the data format is given on the web site. 

\section{Summary}

We reduced and analyzed all absorption line spectra of quasars
taken with the Faint Object Spectrograph on board the Hubble Space
Telescope.  We present line lists and identifications of absorption line
spectra in 271 quasars. Our sample represents a significant increase 
over previous studies; for comparison, the extensive Key Project 
data set (J98), is comprised of spectra for 66 quasars.   
The data set presented here is useful for many studies, particularly those
which benefit from a large number of independent sight-lines. 
We have several studies underway, including 
the study of associated metal-line absorption, 
analysis of the Milky Way ISM lines, and analysis of the proximity effect
to derive the evolution of the UV background at low redshift.  
We have presented the absorption line data in a way which we hope will
enable other workers to carry out additional studies.

%%%%%%%%%%%%%%%%%%%%%%%%%%%%%%%%%%%%%%%%%%%%%%%%%%%%%%%%%%%%%%%%%%%%%%%%

\acknowledgements

We thank Buell Jannuzi for his encouragement and advice.
This research has made use of the NASA/IPAC Extragalactic Database
(NED) which is operated by the Jet Propulsion Laboratory, California
Institute of Technology, under contract with NASA. This project was
supported by STScI grant No.\ AR-05785.02-94A, and
No. GO-066060195A. AD acknowledges support
from NASA Contract No.\ NAS8-39073 (ASC).

%%%%%%%%%%%%%%%%%%%%%%%%%%%%%%%%%%%%%%%%%%%%%%%%%%%%%%%%%%%%%%%%%%%%%%%%
%%% references
%%%%%%%%%%%%%%%%%%%%%%%%%%%%%%%%%%%%%%%%%%%%%%%%%%%%%%%%%%%%%%%%%%%%%%%%

%%%%%%%%%%%%%%%%%%%%%%%%%%%%%%%%%%%%%%%%%%%%%%%%%%%%%%%%%%%%%%%%%%%%%%
% Table 1

\newpage

\begin{deluxetable}{llrrrrll}
\tabletypesize{\scriptsize}
\tablewidth{0pt}
% \tablenum{1}
\tablecaption{Quasars observed with FOS and G130H, G190H or G270H Gratings}
\tablehead{\colhead{Name} & \colhead{Status$^a$} & \colhead{RA} & \colhead{Dec} & \colhead{RA} & \colhead{Dec}& 
\colhead{$z$} & \colhead{Alternate Name}\\
\colhead{}  & \colhead {} & \colhead{(1950)} & \colhead{(1950)} & \colhead{(2000)} & \colhead{(2000)}  & \colhead{} & \colhead{}}
\tablecolumns{8}
\startdata
q0002-4214 &   & 00 02 15.30 & -42 14 11.0 &  00 04 48.20 & -41 57 28.0 &  2.760 & 0002-422  \\  
q0002+0507 &   & 00 02 46.40 & +05 07 29.3 &  00 05 20.20 & +05 24 11.2 & 1.900 &UM18=q0002+051  \\      
q0003+0146 &   & 00 03 13.82 & +01 46 20.4 &  00 05 47.55 & +02 03 02.2 & 0.234& Q0003+0146  \\        
q0003+1553 &   & 00 03 25.14 & +15 53 07.3 &  00 05 59.22 & +16 09 49.1 & 0.450 &0003+15      \\      
q0003+1955 &   & 00 03 45.30 & +19 55 28.5 &  00 06 19.52 & +20 12 10.3 & 0.025 &MRK335       \\      
q0005-2345 &ND & 00 05 27.47 & -23 45 59.6 &  00 08 00.30 & -23 29 18.0 & 1.407& PKS0005-239  \\        
q0007+1041 &   & 00 07 56.74 & +10 41 47.8 &  00 10 30.94 & +10 58 29.0 & 0.089 &IIIZW2       \\         
q0015+1612 &   & 00 15 56.66 & +16 12 46.6 &  00 18 31.90 & +16 29 26.0 & 0.553 & QSO0015+162   \\        
q0017+0209 &   & 00 17 51.15 & +02 09 46.8 &  00 20 25.06 & +02 26 25.3 & 0.401 & Q0017+0209   \\         
q0018+2825 &   & 00 18 01.17 & +28 25 52.4 &  00 20 38.00 & +28 42 31.0 & 0.509 & QSO0020+287   \\        
q0024+2225 &   & 00 24 38.58 & +02 25 23.4 &  00 27 15.40 & +22 41 59.0 & 1.108 &NAB0024+22    \\        
q0026+1259 &   & 00 26 38.13 & +12 59 29.2 &  00 29 13.71 & +13 16 03.8 & 0.142 &PG0026+12     \\       
q0031-7042 &ND & 00 31 58.81 & -70 42 23.6 &  00 34 05.30 & -70 25 52.0 & 0.363 &MC40031-707       \\    
q0038+3242 &ND & 00 08 02.36 & +32 42 05.7 &  00 40 43.50 & +32 58 33.0 & 0.197 &Q0038+327        \\     
q0042+1010 &   & 00 42 02.79 & +10 10 28.9 &  00 44 58.78 & +10 26 52.8 & 0.583 &MC0042+101       \\     
q0043+0354 &   & 00 03 12.60 & +03 54 00.8 &  00 45 47.20 & +04 10 24.0 & 0.384 &PG0043+039       \\     
q0044+0303 &   & 00 04 31.47 & +03 03 32.8 &  00 47 05.90 & +03 19 54.9 & 0.624 &PKS0044+030      \\     
q0050-2523 &ND & 00 50 18.04 & -25 23 09.1 &  00 52 44.70 & -25 06 52.3 & 2.159 &Q0050-253       \\     
q0050+1225 &   & 00 50 57.97 & +12 25 20.2 &  00 53 35.02 & +12 41 36.3 & 0.061 &IZW1            \\      
q0052+2509 &   & 00 52 11.14 & +25 09 24.5 &  00 54 52.13 & +25 25 39.3 & 0.155 &0052+2509     \\        
q0053-0119 &ND & 00 53 33.67 & -01 19 53.7 &  00 56 07.10 & -01 03 40.0 & 0.170 &QSO0056-010   \\        
q0058+0155 &   & 00 58 19.72 & +01 55 28.4 &  01 00 54.10 & +02 11 37.0 & 1.954 &PHL938       \\         
q0059-2735 &ND& 00 59 52.47 & -27 35 56.6 &  01 02 17.01 & -27 19 50.0 & 1.595 &Q0059-2735   \\         
q0100+0205 &   & 01 00 38.61 & +02 05 04.7 &  01 03 12.99 & +02 21 10.4 & 0.394 & 0100+0205   \\          
q0102-2713 &   & 01 02 16.59 & -27 13 12.2 &  01 04 41.00 & -26 57 08.0 & 0.780 &CT336        \\         
q0103-2622 &ND & 01 03 34.43 & -26 22 23.7 &  01 05 59.00 & -26 06 21.0 & 0.776 &Q0103-2622   \\         
q0107-1537 &   & 01 07 03.18 & -15 37 50.2 &  01 09 31.50 & -15 21 52.0 & 0.861 &QSO0107-156  \\         
q0107-0235 &   & 01 07 40.32 & -02 35 51.0 &  01 10 13.15 & -02 19 54.0 & 0.948
&Q0107-025A   \\
q0107-0234 &   & 01 07 43.48 & -02 34 49.7 &  01 10 16.32 & -02 18 52.8 & 0.942 &Q0107-025B   \\         
q0110+2942 &ND & 01 10 38.68 & +29 42 22.9 &  01 13 24.21 & +29 58 15.8 & 0.363 &B20110+29    \\         
q0112-0142 &   & 01 12 44.02 & -01 42 55.1 &  01 15 17.12 & -01 27 04.9 & 1.365 &PKS0112-017  \\         
q0113+3249 &ND & 01 13 19.72 & +32 49 32.4 &  01 16 07.28 & +33 05 21.6 & 0.016 &MRK1         \\         
q0115+0242 &   & 01 15 43.68 & +02 42 19.9 &  01 18 18.46 & +02 58 05.9 & 0.672 &0115+027    \\          
q0117+2118 &   & 01 17 34.69 & +21 18 02.7 &  01 20 17.30 & +21 33 46.0 & 1.493 &PG0117+213  \\          
q0119-0437 &   & 01 19 55.96 & -04 37 08.1 &  01 22 27.90 & -04 21 28.0 & 1.925 &PKS0119-04  \\          
q0121-5903 &   & 01 21 51.24 & -59 03 59.1 &  01 23 45.72 & -58 48 21.8 & 0.047 &FAIRALL9    \\          
q0122-0021 &   & 01 22 55.31 & -00 21 31.6 &  01 25 28.90 & -00 05 56.5 & 1.070 &PKS0122-003  \\         
q0125-0635 &   & 01 25 04.56 & -06 35 07.7 &  01 27 35.47 & -06 19 35.9 & 0.005& SBS0125-065  \\         
q0133+2042 &   & 01 33 40.54 & +20 42 09.4 &  01 36 24.47 & +20 57 26.5 & 0.425 &3C47-0        \\        
q0137+0116 &   & 01 37 22.90 & +01 16 35.7 &  01 39 57.27 & +01 31 46.3 & 0.260 &PHL1093       \\        
q0143-0135 &   & 01 43 18.20 & -01 35 30.8 &  01 45 51.20 & -01 20 31.0 & 3.124 &UM366         \\        
q0150-2015 &   & 01 50 05.04 & -20 15 53.4 &  01 52 27.29 & -20 01 07.2 & 2.139 &UM675         \\        
q0151+0433 &ND & 01 51 51.70 & +04 33 35.6 &  01 54 27.98 & +04 48 17.9 & 0.404 &0151+0433     \\        
q0159-1147 &   & 01 59 30.45 & -11 47 00.0 &  02 01 57.20 & -11 32 33.7 & 0.669 &3C57          \\        
q0200-0858 &ND & 02 00 57.78 & -08 58 11.2 &  02 03 26.18 & -08 43 48.3 & 0.770 &0200-0858     \\        
q0207-3953 &   & 02 07 24.42 & -39 53 49.2 &  02 09 28.60 & -39 39 40.0 & 2.813 &Q0207-398     \\        
q0214+1050 &   & 02 14 26.76 & +10 50 18.4 &  02 17 07.67 & +11 04 09.6 & 0.408 &PKS0214+10    \\
q0219+4248 &   & 02 19 30.06 & +42 48 30.0 &  02 22 39.62 & +43 02 08.1 & 0.444 &3C66A          \\       
q0226-1024 &BAL& 02 26 12.74 & -10 24 32.0 &  02 28 39.15 & -10 11 10.3 & 2.256& Q0226-104     \\        
q0232-0415 &   & 02 32 36.62 & -04 15 10.6 &  02 35 07.30 & -04 02 06.0 & 1.450 &PKS0232-04     \\       
q0237-2322 &ND & 02 37 52.72 & -23 22 08.7 &  02 40 08.10 & -23 09 18.0 & 2.223 &PKS0237-233   \\        
q0240+0044 &ND & 02 40 05.80 & +00 44 18.9 &  02 42 40.12 & +00 57 02.6 & 0.569 &EX0240+007    \\        
q0253-0138 &   & 02 53 44.08 & -01 38 41.2 &  02 56 16.52 & -01 26 37.4 & 0.879 &0253-0138     \\        
q0254-3327 &ND   & 02 54 39.31 & -33 27 27.0 &  02 56 42.49 & -33 15 25.2 & 1.915 &PKS0254-334   \\        
q0254-3327b &  & 02 54 39.42 & -33 27 24.4 &  02 56 42.60 & -33 15 22.6 & 1.915& PKS0254-334  \\         
q0254-3327c &BAL  & 02 54 04.61 & -03 27 29.0 &  02 56 47.78 & -33 15 27.5 & 1.863& Q0254-334    \\         
q0256+3637 &ND & 02 56 49.93 & +36 37 20.7 &  02 59 58.61 & +36 49 14.1 & 0.012& MRK1066      \\         
q0302-2223 &   & 03 02 35.74 & -22 23 29.3 &  03 04 49.80 & -22 11 52.0 & 1.400& 1E0302-223   \\          
q0311+4151 &ND & 03 11 42.86 & +41 51 03.4 &  03 15 01.40 & +42 02 09.4 & 0.024& MRK1073      \\         
q0318-1937 &ND & 03 18 05.56 & -19 37 18.4 &  03 20 21.15 & -19 26 31.7 & 0.104& 0318-196     \\         
q0333+3208 &   & 03 03 02.53 & +32 08 36.3 &  03 36 30.17 & +32 18 29.0 & 1.258& NRAO140     \\          
q0334-3617 &   & 03 34 15.06 & -36 17 25.1 &  03 36 09.28 & -36 07 33.3 & 1.100& 0334-3617   \\          
q0349-1438 &   & 03 49 09.56 & -14 38 06.3 &  03 51 28.60 & -14 29 09.1 & 0.616& 3C95        \\          
q0350-0719 & & 03 50 04.07 & -07 19 05.6 &  03 52 30.56 & -07 11 02.0 & 0.962& 0350-0719   \\          
q0355-4820 &   & 03 55 52.53 & -48 20 48.8 &  03 57 21.90 & -48 12 15.0 & 1.005& 0355-4820    \\         
q0403-1316 &   & 04 03 14.01 & -13 16 18.4 &  04 05 33.98 & -13 08 14.1 & 0.571& PKS0403-13   \\         
q0405-1219 &   & 04 05 27.49 & -12 19 31.7 &  04 07 48.40 & -12 11 36.0 & 0.574& PKS0405-12   \\         
q0414-0601 &   & 04 14 49.33 & -06 01 05.0 &  04 17 16.75 & -05 53 45.9 & 0.781& PKS0414-06=3C110  \\    
q0420-0127 &   & 04 20 43.54 & -01 27 29.7 &  04 23 15.80 & -01 20 34.0 & 0.915& PKS0420-01      \\      
q0421+0157 &   & 04 21 32.71 & +01 57 32.7 &  04 24 08.60 & +02 04 25.0 & 2.044& PKS0421+01     \\      
q0424-1309 &   & 04 24 47.72 & -13 09 33.9 &  04 27 07.20 & -13 02 54.0 & 2.159& PKS0424-13     \\       
q0439-4319 &   & 04 39 42.76 & -43 19 24.4 &  04 41 17.30 & -43 13 43.7 & 0.593& PKS0439-433    \\       
q0450-2958 &   & 04 50 33.03 & -29 58 30.1 &  04 52 30.00 & -29 53 35.0 & 0.286& IR0450-2958    \\       
q0453-4220 &   & 04 53 47.60 & -42 20 59.0 &  04 55 23.00 & -42 16 17.0 & 2.66& Q0453-423       \\       
q0454-2203 &   & 04 54 01.16 & -22 03 49.4 &  04 56 08.90 & -21 59 09.4 & 0.534& PKS0454-22     \\       
q0454+0356 &   & 04 54 09.02 & +03 56 14.3 &  04 56 47.17 & +04 00 52.7 & 1.345& PKS0454+039    \\       
q0506-6113 &   & 05 06 08.59 & -61 13 32.1 &  05 06 44.10 & -61 09 40.0 & 1.093& 0506-6113       \\      
q0513-0012 &ND & 05 13 38.03 & -00 12 15.1 &  05 16 11.51 & -00 08 59.2 & 0.033 &AKN120         \\       
q0518-4549 &   & 05 18 23.60 & -45 49 43.0 &  05 19 49.60 & -45 46 44.0 & 0.0351& PKS0518-45      \\    
q0537-4406 &   & 05 37 21.12 & -44 06 44.8 &  05 38 50.36 & -44 05 09.1 & 0.894 &0537-4406       \\      
q0624+6907 &   & 06 24 35.31 & +69 07 03.2 &  06 30 02.70 & +69 05 04.0 & 0.374 &HS0624+6907   \\       
q0637-7513 &   & 06 37 23.50 & -75 13 37.3 &  06 35 46.70 & -75 16 16.6 & 0.656 &PKS0637-75     \\       
q0710+1151 &   & 07 10 15.41 & +11 51 23.7 &  07 13 02.39 & +11 46 15.5 & 0.768 &3C175=0710+118  \\      
q0740+3800 &   & 07 40 56.82 & +38 00 30.7 &  07 44 17.47 & +37 53 16.9 & 1.063 &3C186=0740+380   \\     
q0742+3150 &   & 07 42 30.83 & +31 50 15.4 &  07 45 41.70 & +31 42 55.7 & 0.462 &B20742+318      \\      
q0743-6719 &   & 07 43 22.28 & -67 19 07.9 &  07 43 31.70 & -67 26 25.0 & 1.511 &PKS0743-67      \\      
q0749+2550 &ND & 07 49 34.82 & +25 50 25.4 &  07 52 37.05 & +25 42 38.5 & 0.446 &OI-287           \\     
q0823-2220 &   & 08 23 50.00 & -22 20 34.6 &  08 26 01.50 & -22 30 27.0 & 0.910 & PKS0823-22       \\     
q0827+2421 &   & 08 27 54.40 & +24 21 07.8 &  08 30 52.09 & +24 10 59.9 & 0.935& B20827+24        \\     
q0830+1133 &ND & 08 30 29.96 & +11 33 52.6 &  08 33 14.40 & +11 23 36.0 & 2.976 &MG0833+112        \\    
q0838+1323 &   & 08 38 01.75 & +13 23 05.8 &  08 40 47.56 & +13 12 23.7 & 0.684 &3C207=0838+133   \\     
q0844+3456 &   & 08 44 33.99 & +34 56 07.9 &  08 47 42.49 & +34 45 03.5 & 0.064 &TON951       \\         
q0848+1623 &   & 08 48 53.67 & +16 23 40.0 &  08 51 41.80 & +16 12 22.0 & 1.936 &Q0848+163      \\       
q0850+4400 &   & 08 50 13.46 & +44 00 24.0 &  08 53 34.20 & +43 49 01.0 & 0.513 &US1867=0850+440  \\
q0851+2017 &   & 08 51 57.33 & +20 17 57.9 &  08 54 48.89 & +20 06 30.1 & 0.306 &OJ287          \\       
q0859-1403 &   & 08 59 54.97 & -14 03 39.2 &  09 02 16.80 & -14 15 31.1 & 1.327 &PKS0859-14      \\      
q0903+1658 &   & 09 03 44.17 & +16 58 16.4 &  09 06 31.86 & +16 46 12.2 & 0.411 &3C215=0903+169  \\      
q0903+1734 & ND & 09 03 50.00 & +17 34 27.6 &  09 06 38.21 & +17 22 23.1 & 2.771 &H0903+175      \\       
q0906+4305 &   & 09 06 17.35 & +43 05 59.6 &  09 09 33.53 & +42 53 47.0 & 0.670 &3C216-0         \\      
q0907-0920 &   & 09 07 04.52 & -09 20 10.1 &  09 09 30.70 & -09 32 24.0 & 0.625 &QSO0909-095-HOPR  \\    
q0916+5118 &   & 09 16 30.10 & +51 18 52.6 &  09 19 57.70 & +51 06 10.0 & 0.553 &NGC2841UB3=0916+513 \\  
q0923+3915 &   & 09 23 55.42 & +39 15 23.7 &  09 27 03.05 & +39 02 20.9 & 0.699 &B20923+39        \\     
q0932+5006 & ND & 09 32 32.15 & +50 06 39.6 &  09 35 53.13 & +49 53 13.6 & 1.920 &Q0932+501        \\     
q0933+7315 &   & 09 33 05.03 & +73 15 27.2 &  09 37 48.80 & +73 01 58.0 & 2.528 &TB0933+733      \\      
q0935+4141 &   & 09 35 48.76 & +41 41 55.5 &  09 38 57.00 & +41 28 21.3 & 1.937 &PG0935+416       \\     
q0940+5425 &ND & 09 40 51.06 & +54 25 14.2 &  09 44 17.22 & +54 11 27.0 & 0.006 &SBS0940+544      \\     
q0945+4053 &   & 09 45 50.13 & +40 53 43.9 &  09 48 55.33 & +40 39 45.0 & 1.252 &4C40-24          \\     
q0946+3009 & ND & 09 46 46.42 & +30 09 20.0 &  09 49 41.11 & +29 55 19.1 & 1.216 &PG0946+301       \\     
q0947+3940 &   & 09 47 44.85 & +39 40 54.4 &  09 50 48.40 & +39 26 51.0 & 0.206 &PG0947+396       \\      
q0953+4129 &   & 09 53 48.26 & +41 29 40.4 &  09 56 52.40 & +41 15 23.0 & 0.239 &PG0953+414       \\     
q0953+5454 &   & 09 53 52.04 & +54 54 34.9 &  09 57 14.70 & +54 40 17.0 & 2.584 &SBS0953+54        \\    
q0954+5537 &   & 09 54 14.39 & +55 37 16.5 &  09 57 38.16 & +55 22 57.7 & 0.909 &4C55-17=0954+556  \\    
q0955+3238 &   & 09 55 25.52 & +32 38 23.1 &  09 58 21.00 & +32 24 02.2 & 0.533 &3C232=0955+326   \\     
q0957+5608a &  & 09 57 57.37 & +56 08 22.1 &  10 01 20.74 & +55 53 55.1 & 1.414 &0957+561A      \\      
q0957+5608b &  & 09 57 57.52 & +56 08 16.2 &  10 01 20.89 & +55 53 49.2 & 1.414& 0957+561B       \\     
q0958+2901 &   & 09 58 57.36 & +29 01 37.7 &  10 01 49.50 & +28 47 09.0 & 0.185 &3C234            \\    
q0958+5509 &   & 09 58 08.23 & +55 09 06.4 &  10 01 29.80 & +54 54 39.0 & 1.75 & MARK132          \\    
q0959+6827 &   & 09 59 09.72 & +68 27 47.8 &  10 03 06.80 & +68 13 17.5 & 0.773& 0959+68W1       \\     
q1001+0527 &   & 10 01 43.30 & +05 27 34.5 &  10 04 20.10 & +05 13 00.0 & 0.161& PG1001+054      \\     
q1001+2239 &   & 10 01 58.54 & +22 39 54.1 &  10 04 45.74 & +22 25 19.1 & 0.974& PKS1001+22    \\       
q1001+2910 &   & 10 01 10.73 & +29 10 08.4 &  10 04 02.60 & +28 55 35.0 & 0.329& TON28=1001+291  \\     
q1007+4147 &   & 10 07 26.12 & +41 47 25.8 &  10 10 27.50 & +41 32 39.1 & 0.611& 4C41.21=1007+417 \\    
q1008+1319 &   & 10 08 29.87 & +13 19 00.5 &  10 11 10.80 & +13 04 12.0 & 1.287& PG1008+133       \\    
q1010+3606 &   & 10 10 07.39 & +36 06 15.1 &  10 13 03.17 & +35 51 23.1 & 0.070& CSO251          \\     
q1017+2759 &   & 10 17 07.82 & +27 59 06.7 &  10 19 56.63 & +27 44 01.3 & 1.928& TON34           \\     
q1026-0045a &  & 10 26 01.66 & -00 45 22.6 &  10 28 35.00 & -01 00 44.0 & 1.437& Q1026-0045-A    \\     
q1026-0045b &  & 10 26 03.65 & -00 45 06.6 &  10 28 37.00 & -01 00 28.0 & 1.53 & Q1026-0045-B    \\     
q1028+3118 &   & 10 28 09.81 & +31 18 21.1 &  10 30 59.10 & +31 02 56.0 & 0.178 &B21028+313      \\     
q1030+6017 & ND & 10 30 51.49 & +60 17 21.9 &  10 34 08.57 & +60 01 52.0 & 0.051& MARK34         \\      
q1038+0625 &   & 10 38 40.99 & +06 25 58.6 &  10 41 17.20 & +06 10 17.0 & 1.270& 4C06.41        \\      
q1047+5503 &   & 10 47 43.15 & +55 03 13.6 &  10 50 45.80 & +54 47 19.0 & 2.165& FBS1047+550    \\      
q1049-0035 &   & 10 49 18.06 & -00 35 21.7 &  10 51 51.50 & -00 51 18.1 & 0.357& PG1049-005     \\      
q1055+2007 &   & 10 55 37.61 & +20 07 05.1 &  10 58 17.90 & +19 51 50.7 & 1.110& PKS1055+20     \\      
q1100-2629 &   & 11 00 59.93 & -26 29 05.4 &  11 03 25.26 & -26 45 15.7 & 2.145& Q1101-264      \\      
q1100+7715 &   & 11 00 27.51 & +77 15 08.7 &  11 04 13.90 & +76 58 58.2 & 0.311& 3C249-1=1100+772 \\    
q1101+3828 & ND & 11 01 40.64 & +38 28 42.8 &  11 04 27.32 & +38 12 31.5 & 0.031& MRK421           \\    
q1103-0036 &   & 11 03 58.27 & -00 36 39.8 &  11 06 31.75 & -00 52 53.4 & 0.425& PKS1103-006     \\     
q1103+6416 &   & 11 03 04.08 & +64 16 21.9 &  11 06 10.80 & +64 00 09.0 & 2.19 & HS-1103+6416     \\    
q1104-1805a &  & 11 04 04.95 & -18 05 10.1 &  11 06 33.50 & -18 21 24.0 & 2.303& HE-1104-1805A    \\    
q1104-1805b &  & 11 04 05.05 & -18 05 12.1 &  11 06 33.60 & -18 21 26.0 & 2.303& HE-1104-1805B   \\     
q1104+1644 &   & 11 04 36.66 & +16 44 16.7 &  11 07 15.00 & +16 28 02.4 & 0.634& MC1104+167           \\
q1111+4053 &   & 11 11 53.35 & +40 53 41.4 &  11 14 38.70 & +40 37 20.1 & 0.734& 3C254=1111+408   \\    
q1114+4429 &   & 11 14 20.00 & +44 29 57.4 &  11 17 06.30 & +44 13 34.0 & 0.144 &PG1114+445      \\     
q1115+0802a1 &  &  11 15 41.52 & +08 02 23.3 & 11 18 16.94 & +07 45 58.9 & 1.722& PG1115+080A1   \\     
q1115+0802a2 &  & 11 15 41.52 & +08 02 23.3 & 11 18 16.94 & +07 45 58.9 & 1.722& PG1115+080A2    \\    
q1115+4042 &   & 11 15 45.91 & +40 42 19.6 & 11 18 30.30 & +40 25 55.0 & 0.154&  PG1115+407      \\    
q1116+2135 &   & 11 16 30.21 & +21 35 43.1 & 11 19 08.70 & +21 19 18.0 & 0.117& PG1116+215       \\    
q1118+1252 &   & 11 18 53.46 & +12 52 43.8 & 11 21 29.76 & +12 36 16.9 & 0.685& MC1118+12        \\    
q1120+0154 & ND & 11 20 46.67 & +01 54 16.3 & 11 23 20.70 & +01 37 48.0 & 1.465& UM425           \\     
q1121+4218 &   & 11 21 55.83 & +42 18 14.4 & 11 24 39.17 & +42 01 45.2 & 0.234& Q1121+423       \\     
q1122-1648 &   & 11 22 12.28 & -16 48 48.4 & 11 24 42.80 & -17 05 18.0 & 2.40 & HE-1122-1649    \\     
q1123+3935 & ND & 11 23 45.98 & +39 35 15.4 & 11 26 28.00 & +39 18 45.0 & 1.470& B31123+395      \\     
q1124+2711 &   & 11 24 57.64 & +27 11 21.3 & 11 27 36.40 & +26 54 50.0 & 0.378& QSO1127+269     \\     
q1127-1432 &   & 11 27 35.71 & -14 32 54.8 & 11 30 07.02 & -14 49 27.6 & 1.187& PKS1127-14      \\     
q1129-0229 & ND& 11 29 56.55 & -02 29 25.9 & 11 32 29.85 & -02 46 00.0 & 0.333& Q1129-0229      \\     
q1130+1108 &   & 11 30 55.04 & +11 08 57.7 & 11 33 30.30 & +10 52 23.0 & 0.510& 1130+106Y=1130+111 \\  
q1132-0302 &   & 11 32 31.64 & -03 02 17.0 & 11 35 04.90 & -03 18 52.5 & 0.237& Q1132-0302      \\     
q1136-1334 &   & 11 36 38.58 & -13 34 05.7 & 11 39 10.70 & -13 50 43.1 & 0.557& PKS1136-135     \\     
q1137+6604 &   & 11 37 09.44 & +66 04 27.0 & 11 39 57.10 & +65 47 49.4 & 0.652& 3C263.0=1137+660 \\    
q1138+0204 &   & 11 38 47.83 & +02 04 41.5 & 11 41 21.71 & +01 48 03.3 & 0.383& Q1138+0204    \\       
q1144-0115 &   & 11 44 44.44 & -01 15 27.5 & 11 47 18.03 & -01 32 07.7 & 0.382& Q1144-0115=1144-012 \\ 
q1144-3755 &ND & 11 44 30.88 & -37 55 31.0 & 11 47 01.32 & -38 12 11.1 & 1.048& 1144-3755      \\      
q1146+1106c &  &  11 46 04.93 & +11 06 57.7 & 11 48 39.40 & +10 50 17.0 & 1.01 &
 1146+111C     \\
q1146+1103e &  &  11 46 07.54 & +11 03 49.7 & 11 48 42.00 & +10 47 09.0 & 1.10 & 1146+111E      \\      
q1146+1104b &  &  11 46 09.84 & +11 04 37.7 & 11 48 44.30 & +10 47 57.0 & 1.01 & 1146+111B      \\      
q1146+1105d &ND &  11 46 16.84 & +11 05 04.7 & 11 48 51.30 & +10 48 24.0 & 2.12 & 1146+111D     \\       
q1146+1111 & ND & 11 46 13.46 & +11 11 39.1 & 11 48 47.86 & +10 54 58.6 & 0.863& MC1146+111   \\        
q1148+5454 &   & 11 48 42.64 & +54 54 13.9 & 11 51 20.44 & +54 37 32.8 & 0.978& 1148+5454     \\       
q1150+4947 &   & 11 50 48.06 & +49 47 50.0 & 11 53 24.46 & +49 31 08.6 & 0.334& LB2136         \\      
q1156+2123 &   & 11 56 52.30 & +21 23 38.2 & 11 59 26.20 & +21 06 56.2 & 0.349& TEX1156+213    \\      
q1156+2931 &   & 11 56 57.93 & +29 31 25.6 & 11 59 31.90 & +29 14 43.6 & 0.729& 4C29.45=1156+295 \\    
q1156+6311 &   & 11 56 04.72 & +63 11 10.0 & 11 58 39.91 & +62 54 28.1 & 0.594& 1156+6311    \\        
q1202+2810 &   & 12 02 09.05 & +28 10 53.9 & 12 04 42.20 & +27 54 12.0 & 0.165& PG1202+281   \\        
q1206+4557 &   & 12 06 26.62 & +45 57 17.4 & 12 08 58.00 & +45 40 36.0 & 1.158& PG1206+459    \\       
q1211+1419 &   & 12 11 44.95 & +14 19 52.7 & 12 14 17.68 & +14 03 12.3 & 0.085& PG1211+1431   \\       
q1213-0017 & ND &  12 13 16.01 & -00 17 53.8 & 12 15 49.80 & -00 34 34.0 & 2.69 & UM485         \\       
q1214-2745 &ND & 12 14 40.91 & -27 45 52.8 & 12 17 17.06 & -28 02 32.4 & 0.026& T1214-277     \\       
q1214+1804 &   & 12 14 16.83 & +18 04 44.2 & 12 16 49.05 & +17 48 04.5 & 0.375& Q1214+1804    \\       
q1215+6423 &   & 12 15 17.13 & +64 23 46.8 & 12 17 40.86 & +64 07 07.4 & 1.288& 4C64-15       \\       
q1216-0102 &ND & 12 16 09.13 & -01 02 47.5 & 12 18 42.92 & -01 19 26.6 & 0.415& 1216-0103     \\       
q1216+0655 &   & 12 16 47.84 & +06 55 17.3 & 12 19 20.90 & +06 38 38.4 & 0.334& PG1216+069     \\      
q1216+5032a&   &  12 16 13.49 & +50 32 15.1 & 12 18 41.10 & +50 15 36.0  & 1.450& HS-1216+5032A  \\      
q1216+5032b & ND &  12 16 12.99 & +50 32 23.1 & 12 18 40.60 & +50 15 44.0 & 1.451& HS-1216+5032B \\	 
q1219+0447 &   & 12 19 04.73 & +04 47 03.8 & 12 21 37.94 & +04 30 25.7 & 0.094& 1219+047       \\      
q1219+7535 &   & 12 19 33.51 & +75 35 16.1 & 12 21 44.04 & +75 18 38.1 & 0.070& 1219+7535 MRK 205 \\   
q1220+1601 &   & 12 20 58.98 & +16 01 45.2 & 12 23 30.82 & +15 45 07.9 & 0.081& 1220+1601      \\      
q1222+2251 &   & 12 22 56.65 & +22 51 49.4 & 12 25 27.39 & +22 35 13.0 & 2.046& PG1222+228     \\      
q1224-1116 & ND & 12 24 47.18 & -11 16 58.9 & 12 27 22.40 & -11 33 34.4 & 1.979& Q1224-1116           \\
q1225+3145 &   & 12 25 56.10 & +31 45 12.1 & 12 28 24.92 & +31 28 37.1 & 2.219& B21225+317     \\      
q1226+0219 &   & 12 26 33.25 & +02 19 43.46 & 12 29 06.70 & +02 03 08.60 &0.158&
 3C273  \\
q1229-0207 &   & 12 29 25.96 & -02 07 32.4 & 12 32 00.00 & -02 24 05.4 & 1.045& PKS1229-02     \\      
q1229+6430 &ND & 12 29 16.13 & +64 30 50.8 & 12 31 31.40 & +64 14 17.6 & 0.170& 1229+6430      \\   
q1230+0947 &   & 12 30 53.69 & +09 47 55.2 & 12 33 25.79 & +09 31 23.0 & 0.420& 1230+0947       \\     
q1241+1737 &   & 12 41 41.03 & +17 37 28.5 & 12 44 10.80 & +17 21 04.0 & 1.273& PG1241+176    \\       
q1244+3225 &ND & 12 44 55.48 & +32 25 28.8 & 12 47 20.80 & +32 09 07.0 & 0.949& B21244+32B     \\      
q1245+3431 &ND & 12 45 03.21 & +34 31 30.7 & 12 47 27.80 & +34 15 09.0 & 2.068 &B19             \\     
q1246-0542 & ND & 12 46 38.73 & -05 42 59.1 & 12 49 13.84 & -05 59 19.3 & 2.236& 1246-057       \\      
q1247+2647 &   & 12 47 39.14 & +26 47 26.8 & 12 50 05.73 & +26 31 07.5 & 2.043& PG1247+267     \\      
q1248+3032 &   & 12 48 00.21 & +30 32 58.8 & 12 50 25.54 & +30 16 39.8 & 1.061& B21248+30     \\       
q1248+3142 &   & 12 48 25.36 & +31 42 11.6 & 12 50 50.30 & +31 25 53.0 & 1.02 & CSO173          \\     
q1248+4007 &   & 12 48 26.65 & +40 07 58.2 & 12 50 48.30 & +39 51 39.6 & 1.030& PG1248+401     \\      
q1249+2929 &   & 12 49 59.57 & +29 29 38.2 & 12 52 25.00 & +29 13 21.0 & 0.82 & CSO176         \\      
q1250+3122 &   & 12 50 52.87 & +31 22 06.3 & 12 53 17.50 & +31 05 50.0 & 0.78 & CSO179         \\      
q1250+5650 &   & 12 50 15.20 & +56 50 37.4 & 12 52 26.29 & +56 34 20.4 & 0.321& 3C277-1=1250+568 \\    
q1252+1157 &   & 12 52 07.74 & +11 57 21.1 & 12 54 38.20 & +11 41 06.1 & 0.871& PKS1252+11     \\      
q1253-0531 &   & 12 53 35.89 & -05 31 08.3 & 12 56 01.15 & -05 47 21.7 & 0.538& 3C279=1253-055  \\     
q1254+0443 & ND & 12 54 27.51 & +04 43 46.6 & 12 56 59.90 & +04 27 34.1 & 1.024& PG1254+047      \\     
q1254+5708 &ND & 12 54 05.04 & +57 08 37.1 & 12 56 14.20 & +56 52 24.0 & 0.042& MRK231          \\     
q1257+2840 &ND & 12 57 57.79 & +28 40 10.6 & 13 00 22.20 & +28 24 01.8 & 0.092& 1257+2840       \\     
q1257+3439 &   & 12 57 26.64 & +34 39 31.4 & 12 59 48.70 & +34 23 22.0 & 1.375& B2011257+34=1257+346 \\  
q1258+2835 &   & 12 58 36.56 & +28 35 52.3 & 13 01 00.90 & +28 19 44.2 & 1.355& QSO-1258+285    \\     
q1259+5918 &   & 12 59 08.26 & +59 18 14.6 & 13 01 12.90 & +59 02 06.9 & 0.472& PG1259+593      \\     
q1302-1017 &   & 13 02 55.91 & -10 17 17.5 & 13 05 33.00 & -10 33 20.4 & 0.286& PKS1302-102      \\    
q1305+0658 &   & 13 05 22.57 & +06 58 14.2 & 13 07 53.96 & +06 42 14.2 & 0.602& 3C281=1305+069  \\     
q1306+3021 &   & 13 06 07.27 & +30 21 38.1 & 13 08 29.69 & +30 05 39.0 & 0.806& 1306+3021        \\    
q1307+0835 &   & 13 07 16.27 & +08 35 47.4 & 13 09 47.03 & +08 19 49.8 & 0.155& 1307+0835         \\   
q1307+4617 &   & 13 07 58.59 & +46 17 20.8 & 13 10 11.70 & +46 01 24.0 & 2.129& HS-1307+4617    \\     
q1309-0536 & ND& 13 09 00.76 & -05 36 43.7 & 13 11 36.48 & -05 52 38.9 & 2.188 & Q1309-0536      \\     
q1309+3531 &   & 13 09 58.43 & +35 31 15.2 & 13 12 17.70 & +35 15 21.0 & 0.184 &PG1309+355        \\   
q1311+0217 &   & 13 11 53.60 & +02 17 05.7 & 13 14 26.50 & +02 01 14.3 & 0.306& Q1311+0217       \\    
q1317-0142 &   & 13 17 15.89 & -01 42 20.9 & 13 19 50.30 & -01 58 04.6 & 0.225& Q1317-0142      \\     
q1317+2743 &   & 13 17 34.38 & +27 43 51.8 & 13 19 56.30 & +27 28 08.4 & 1.022& TON153=1317+277  \\    
q1317+5203 &   & 13 17 41.27 & +52 03 49.6 & 13 19 46.23 & +51 48 06.1 & 1.055& 1317+5203       \\     
q1318+2903 &   & 13 18 54.67 & +29 03 01.5 & 13 21 15.80 & +28 47 20.0 & 0.549 &TON156           \\    
q1318+2903a&   & 13 18 53.58 & +29 03 30.5 & 13 21 14.70 & +28 47 49.0 & 1.703 &TON155           \\    
q1320+2925 &   & 13 20 59.89 & +29 25 45.3 & 13 23 20.50 & +29 10 07.0 & 0.960 &TON157           \\    
q1321+0552 &   & 13 21 48.52 & +05 52 42.1 & 13 24 19.90 & +05 37 05.0 & 0.205 &IR1321+0552     \\     
q1322+6557 &   & 13 22 08.46 & +65 57 25.0 & 13 23 49.50 & +65 41 48.0 & 0.168 &PG1322+659       \\    
q1323+6530 &   & 13 23 48.58 & +65 30 47.3 & 13 25 29.70 & +65 15 13.0 & 1.618 &4C65.15         \\     
q1327-2040 &   & 13 27 24.34 & -20 40 48.9 & 13 30 07.70 & -20 56 17.0 & 1.169 &PKS-1327-206-FIX \\    
q1328+3045 &   & 13 28 49.74 & +30 45 58.6 & 13 31 08.29 & +30 30 32.8 & 0.849& 3C286.0         \\     
q1329+4117 &   & 13 29 29.82 & +41 17 23.7 & 13 31 41.10 & +41 01 59.0 & 1.93 & PG1329+412       \\    
q1333+1740 &   & 13 33 36.81 & +17 40 30.5 & 13 36 02.00 & +17 25 13.0 & 0.554& PG1333+176     \\      
q1334-0033 &   & 13 34 13.12 & -00 33 41.4 & 13 36 47.13 & -00 48 57.7 & 2.783& QSO-133647-00485  \\   
q1334+2438 &  &  13 34 57.34 & +24 38 18.2 & 13 37 18.70 & +24 23 03.0 & 0.108& IR1334+24       \\     
q1338+4138 &  &  13 38 52.06 & +41 38 22.3 & 13 41 00.77 & +41 23 14.1 & 1.219& 1338+416            \\
q1340-0038 &  &  13 40 17.51 & -00 38 40.8 & 13 42 51.58 & -00 53 46.0 & 0.326& Q1340-0038      \\     
q1340+6036 &  &  13 40 30.06 & +60 36 48.4 & 13 42 13.24 & +60 21 42.8 & 0.961& 3C288-1         \\     
q1346+2637 &  &  13 46 29.61 & +26 37 13.5 & 13 48 48.30 & +26 22 20.0 & 0.598 &QSO1348+263      \\    
q1351+3153 &  &  13 51 51.28 & +31 53 45.1 & 13 54 05.39 & +31 39 02.4 & 1.326& B21351+31       \\     
q1351+6400 &  &  13 51 46.42 & +64 00 29.3 & 13 53 15.86 & +63 45 45.6 & 0.088& PG1351+64         \\   
q1352+0106 &  &  13 52 25.56 & +01 06 51.2 & 13 54 58.70 & +00 52 10.0 & 1.117& PG1352+011       \\    
q1352+1819 &  &  13 52 12.60 & +18 19 58.9 & 13 54 35.70 & +18 05 17.0 & 0.152& PG1352+183        \\   
q1354+1933 &  &  13 54 42.23 & +19 33 43.2 & 13 57 04.50 & +19 19 06.6 & 0.719& PKS1354+19       \\    
q1354+2552 & ND & 13 54 48.35 & +25 52 01.5 & 13 57 06.50 & +25 37 25.0 & 2.032& PKS1354+25       \\    
q1355-4138 &  &  13 55 57.24 & -41 38 20.3 & 13 59 00.18 & -41 52 53.5 & 0.313& 1355-413         \\    
q1356+5806 &  &  13 56 36.31 & +58 06 37.7 & 13 58 17.60 & +57 52 04.4 & 1.375& 4C58-29         \\     
q1401+0952 &  &  14 01 42.78 & +09 52 06.8 & 14 04 10.60 & +09 37 45.5 & 0.441& 1401+0951         \\   
q1402+4341 &ND &  14 02 37.75 & +43 41 27.0 & 14 04 38.79 & +43 27 07.3 & 0.320&
 Q1402+4341       \\
q1402+2609 &  &  14 02 59.34 & +26 09 52.2 & 14 05 16.20 & +25 55 33.6 & 0.164& 1402+2609        \\    
q1404+2238 &  &  14 04 02.52 & +22 38 03.2 & 14 06 21.90 & +22 23 47.0 & 0.098 &PG1404+226       \\    
q1407+2632 &  &  14 07 07.82 & +26 32 30.3 & 14 09 23.88 & +26 18 21.2 & 0.944& PG1407+265      \\     
q1408+5642 &ND&  14 08 14.60 & +56 42 36.2 & 14 09 54.21 & +56 28 28.9 & 2.562& Q1408+5642       \\    
q1411+4414 &ND&  14 11 50.09 & +44 14 11.4 & 14 13 48.30 & +44 00 13.0 & 0.089 &PG1411+442       \\    
q1413+1143 &ND & 14 13 20.22 & +11 43 38.3 & 14 15 46.30 & +11 29 44.1 & 2.551& H1413+117        \\    
q1415+4343 &  & 14 15 03.31 & +43 43 55.5 & 14 17 01.39 & +43 30 04.9 & 0.002& SBS1415+437      \\    
q1415+4509 &  &  14 15 04.65 & +45 09 56.6 & 14 17 00.91 & +44 56 06.0 & 0.114 &PG1415+451        \\   
q1416-1256 &  &  14 16 21.36 & -12 56 58.3 & 14 19 03.90 & -13 10 45.0 & 0.129 &PG1416-129        \\   
q1416+0642 &  &  14 16 38.80 & +06 42 20.2 & 14 19 08.20 & +06 28 34.0 & 1.436 &3C298            \\    
q1424-1150 &  &  14 24 55.99 & -11 50 25.2 & 14 27 38.10 & -12 03 49.9 & 0.806 &PKS1424-118      \\    
q1425+2003 &  &  14 25 05.80 & +20 03 17.1 & 14 27 25.05 & +19 49 52.3 & 0.111 &1425+2003        \\    
q1425+2645 &  &  14 25 21.93 & +26 45 39.3 & 14 27 35.70 & +26 32 15.0 & 0.366 &B21425+26         \\   
q1427+4800 &  &  14 27 54.04 & +48 00 44.1 & 14 29 43.10 & +47 47 26.0 & 0.221 &PG1427+480        \\   
q1435-0134 &  &  14 35 13.29 & -01 34 13.5 & 14 37 48.30 & -01 47 11.0 & 1.310 &Q1435-0134        \\   
q1435+6349 &  &  14 35 37.30 & +63 49 36.3 & 14 36 45.77 & +63 36 38.1 & 2.068 &S4-1435+638      \\    
q1439+5343 &ND&  14 39 02.59 & +53 43 03.7 & 14 40 38.06 & +53 30 15.7 & 0.038 &MRK477           \\    
q1440+3539 &  &  14 40 04.63 & +35 39 07.4 & 14 42 07.46 & +35 26 22.9 & 0.077 &1440+3539  MARK-478 \\  
q1444+4047 &  &  14 44 50.24 & +40 47 38.1 & 14 46 45.91 & +40 35 07.1 & 0.267 &PG1444+407       \\    
q1451-3735 &  &  14 51 18.33 & -37 35 24.0 & 14 54 27.36 & -37 47 34.2 & 0.314 &PKS1451-375       \\   
q1503+6907 &  &  15 03 44.32 & +69 07 49.0 & 15 04 12.70 & +68 56 13.0 & 0.318 &B2-1503+691       \\   
q1512+3701 &  &  15 12 47.42 & +37 01 55.7 & 15 14 43.52 & +36 50 51.0 & 0.371 &B2-1512+37       \\    
q1517+2356 &  &  15 17 08.26 & +23 56 53.3 & 15 19 19.46 & +23 46 03.3 & 1.903 &LB9612          \\     
q1517+2357 &  &  15 17 02.19 & +23 57 44.7 & 15 19 13.39 & +23 46 54.4 & 1.834 &LB9605           \\    
q1521+1009 &  &  15 21 59.93 & +10 09 02.9 & 15 24 24.52 & + 9 58 29.5 & 1.324 &PG1522+101        \\   
q1524+5147 & ND & 15 24 26.04 & +51 47 15.5 & 15 25 53.80 & +51 36 49.0 & 2.860 &1524+5147       \\     
q1535+5443& ND& 15 35 20.75 & +54 43 22.0 & 15 36 38.36 & +54 33 33.1 & 0.039& MRK486         \\      
q1538+4745 &  &  15 38 00.96 & +47 45 10.6 & 15 39 34.79 & +47 35 31.6 & 0.770 &PG1538+477      \\     
q1540+1805 &ND&  15 40 03.58 & +18 05 37.5 & 15 42 19.50 & +17 56 07.0 & 1.662 &4C18.43          \\    
q1542+5408 &  &  15 42 41.88 & +54 08 25.6 & 15 43 59.40 & +53 59 03.0  & 2.371 &SBS1542+541      \\    
q1544+4855 &  &  15 44 00.16 & +48 55 25.5 & 15 45 30.34 & +48 46 07.9 & 0.400 &1543+4855        \\    
q1545+2101 &  &  15 45 31.21 & +21 01 27.2 & 15 47 43.53 & +20 52 16.4 & 0.264 &3C323-1=1545+210 \\    
q1555+3313 &  &  15 55 33.72 & +33 13 20.8 & 15 57 29.95 & +33 04 46.7 & 0.942 &B21555+33        \\    
q1611+3420 &  &  16 11 48.05 & +34 20 20.0 & 16 13 41.10 & +34 12 48.0 & 1.401 &DA406=1611+343       \\
q1612+2611 &  &  16 12 08.82 & +26 11 46.5 & 16 14 13.23 & +26 04 16.2 & 0.131& 1612+2611      \\      
q1615+3229 &  &  16 15 46.90 & +32 29 51.0 & 16 17 42.50 & +32 22 25.0 & 0.151 & 3C 332 \\
q1618+1743 &  &  16 18 07.37 & +17 43 30.5 & 16 20 21.80 & +17 36 24.0 & 0.555 &3C334.0\\    
q1622+2352 &  &  16 22 32.30 & +23 52 01.3 & 16 24 39.08 & +23 45 12.1 & 0.927 &3C336.0          \\    
q1623+2653 &  &  16 23 46.05 & +26 53 43.4 & 16 25 48.90 & +26 46 59.0 &  2.526 &1623.7+268B      \\    
q1626+5529 &  &  16 26 51.47 & +55 29 04.9 & 16 27 56.10 & +55 22 31.0 & 0.133 &PG1626+554       \\    
q1630+3744 &  &  16 30 15.18 & +37 44 08.3 & 16 32 01.10 & +37 37 49.4 & 1.478 &1630+377          \\   
q1631+3930 &  &  16 31 19.41 & +39 30 41.8 & 16 33 02.10 & +39 24 27.2 & 1.023 &1631+3930        \\    
q1634+7037 &  &  16 34 51.83 & +70 37 37.2 & 16 34 29.05 & +70 31 32.7 & 1.337 &PG1634+706       \\    
q1637+5726 &  &  16 37 17.53 & +57 26 15.8 & 16 38 13.46 & +57 20 23.9 & 0.745 &OS562=1637+574   \\    
q1641+3954 &  &  16 41 17.67 & +39 54 10.8 & 16 42 58.78 & +39 48 36.9 & 0.595 &3C345=1641+399    \\   
q1652+3950 &ND&  16 52 11.85 & +39 50 25.2 & 16 53 52.24 & +39 45 36.6 & 0.033 &MRK501           \\    
q1656+0519 &  &  16 56 05.73 & +05 19 47.1 & 16 58 33.46 & +05 15 16.5 & 0.887 &PKS1656+053      \\    
q1700+5153 & ND & 17 00 13.46 & +51 53 35.9 & 17 01 24.90 & +51 49 20.0 & 0.288 &PG1700+518       \\    
q1700+6416 &  &  17 00 40.63 & +64 16 24.7 & 17 01 00.48 & +64 12 08.9 & 2.722 &HS1700+6416     \\     
q1701+6102 &ND & 17 01 34.15 & +61 02 59.4 & 17 02 11.11 & +60 58 48.0 & 0.164& 1701+6102       \\     
q1704+6048 &  &  17 04 03.52 & +60 48 30.8 & 17 04 41.30 & +60 44 30.0 & 0.371 &3C351.0=1704+608  \\   
q1715+5331 &  &  17 15 30.33 & +53 31 27.2 & 17 16 35.40 & +53 28 16.2 & 1.940 &PG1715+535      \\     
q1717+4901 &  &  17 17 56.45 & +49 01 49.7 & 17 19 14.53 & +48 58 49.6 & 0.025 &ARP102B         \\     
q1718+4807 &  &  17 18 17.86 & +48 07 10.6 & 17 19 38.30 & +48 04 12.2 & 1.084 &PG1718+481       \\    
q1803+7827 &  &  18 03 39.36 & +78 27 54.4 & 18 00 45.70 & +78 28 04.1 & 0.680 &S51803+78         \\   
q1821+1042 &  &  18 21 41.71 & +10 42 45.1 & 18 24 02.90 & +10 44 25.0 & 1.360 &PKS1821+10         \\  
q1821+6419 &  &  18 21 44.13 & +64 19 32.1 & 18 21 59.40 & +64 21 07.5 & 0.297 &E1821+643      \\      
q1845+7943 &  &  18 45 37.49 & +79 43 06.2 & 18 42 08.80 & +79 46 17.0 & 0.056 &3C390.3          \\    
q1928+7351 &  &  19 28 49.49 & +73 51 45.3 & 19 27 48.50 & +73 58 02.0 & 0.302 &4C73.18=1928+738 \\    
q1935-6914 &ND&  19 35 11.88 & -69 14 51.3 & 19 40 25.56 & -69 07 56.5 & 3.152 &QSO-194025-69075 \\    
q2041-1054 &  &  20 41 26.40 & -10 54 17.9 & 20 44 09.76 & -10 43 24.5 & 0.035 &MRK509          \\     
q2112+0555 &  &  21 12 23.47 & +05 55 12.9 & 21 14 52.57 & +06 07 42.5 & 0.457 &PG2112+059     \\      
q2128-1220 &  &  21 28 52.93 & -12 20 21.1 & 21 31 35.42 & -12 07 05.5 & 0.501 &PKS2128-12      \\     
q2135-1446 &  &  21 35 01.25 & -14 46 27.4 & 21 37 45.19 & -14 32 55.8 & 0.200 &PKS2135-147     \\     
q2141+1730 &  &  21 41 13.86 & +17 30 02.3 & 21 43 35.56 & +17 43 49.1 & 0.213 &OX1692141+175   \\     
q2145+0643 &  &  21 45 36.21 & +06 43 41.4 & 21 48 05.50 & +06 57 39.0 & 0.999 &PKS2145+067     \\     
q2155-3027 &  &  21 55 58.38 & -30 27 54.4 & 21 58 52.00 & -30 13 32.3 & 0.116 &PKS2155-304     \\     
q2200+4202 &ND&  22 00 39.47 & +42 02 08.4 & 22 02 43.30 & +42 16 39.8 & 0.069 &BL-LAC         \\      
q2201+3131 &  &  22 01 01.58 & +31 31 05.3 & 22 03 15.02 & +31 45 37.7 & 0.297 &B22201+31A      \\     
q2212-2959 &  &  22 12 25.08 & -29 59 20.8 & 22 15 16.00 & -29 44 24.0 & 2.703 &PKS2212-299     \\     
q2215-0347 &ND&  22 15 12.21 & -03 47 40.6 & 22 17 47.77 & -03 32 39.0 & 0.241 &2215-0347      \\      
q2216-0350 &  &  22 16 16.52 & -03 50 41.0 & 22 18 52.08 & -03 35 37.4 & 0.901 &PKS2216-03      \\     
q2223-0512 &  &  22 23 11.22 & -05 12 17.7 & 22 25 47.31 & -04 57 01.4 & 1.404 &3C446=2223-052  \\     
q2230+1128 &  &  22 30 07.94 & +11 28 22.7 & 22 32 36.45 & +11 43 50.7 & 1.037 &CTA102=2230+114  \\    
q2243-1222 &  &  22 43 39.85 & -12 22 40.0 & 22 46 18.20 & -12 06 51.2 & 0.630 &PKS2243-123     \\     
q2251-1750 &  &  22 51 25.85 & -17 50 54.6 & 22 54 05.80 & -17 34 55.0 & 0.068 &MR2251-178       \\    
q2251+1120 &  &  22 51 40.67 & +11 20 39.5 & 22 54 10.44 & +11 36 38.9 & 0.323 &PKS2251+113     \\     
q2251+1552 &  &  22 51 29.66 & +15 52 54.2 & 22 53 57.80 & +16 08 53.4 & 0.859 &3C454.3=2251+158 \\    
q2300-6823 &  &  23 00 27.93 & -68 23 47.3 & 23 03 43.50 & -68 07 37.1 & 0.512 &PKS2300-683      \\    
q2302+0255 &  &  23 02 12.10 & +02 55 34.0 & 23 04 44.91 & +03 11 45.7 & 1.052& PG2302+029     \\      
q2308+0951 &  &  23 08 46.67 & +09 51 57.9 & 23 11 17.81 & +10 08 16.2 & 0.432 &PG2308+098       \\  
q2340-0339 &  &  23 40 22.59 & -03 39 05.3 & 23 42 56.60 & -03 22 26.5 & 0.896 &PKS2340-036      \\    
q2344+0914 &  &  23 44 03.92 & +09 14 06.0 & 23 46 36.90 & +09 30 46.0 & 0.672 &PKS2344+092      \\    
q2347-4342 &  &  23 47 57.40 & -43 42 41.0 & 23 50 34.20 & -43 26 00.0 & 2.90  &HE-2347-4342     \\    
q2349-0125 &  &  23 49 22.28 & -01 25 54.9 & 23 51 56.03 & -01 09 13.7 & 0.174 &2349-0125       \\     
q2352-3414 &  &  23 52 50.76 & -34 14 37.5 & 23 55 25.60 & -33 57 55.8 & 0.706 &PKS2352-342     \\     
\enddata
\tablenotetext{a}{ND: data for this object not analyzed  -- 
observation failed, the spectrum contained 
no signal, data was obtained in SPECTROPOLARIMETRY mode,  or the object is
a broad absorption line quasar.}
\end{deluxetable}

%%%%%%%%%%%%%%%%%%%%%%%%%%%%%%%%%%%%%%%%%%%%%%%%%%%%%%%%%%%%%%%%%%%%%%%%
%%% table2 and 3
%%%%%%%%%%%%%%%%%%%%%%%%%%%%%%%%%%%%%%%%%%%%%%%%%%%%%%%%%%%%%%%%%%%%%%%%

\clearpage

\begin{center}
Table 2. Observation Details.
\end{center}

Available in:

\verb@      http://lithops.as.arizona.edu/~jill/QuasarSpectra/@

or:

\verb@      http://hea-www.harvard.edu/QEDT/QuasarSpectra/@.

\vspace*{1in}

\begin{center}

Table 3. Absorption line lists and line identifications.

\end{center}

Available in:

\verb@      http://lithops.as.arizona.edu/~jill/QuasarSpectra/@

or:

\verb@      http://hea-www.harvard.edu/QEDT/QuasarSpectra/@.

\vspace*{1in}

\begin{center}
Table 4. Objects which were only observed Pre-Costar in A-1.
\end{center}

Available in:

\verb@      http://lithops.as.arizona.edu/~jill/QuasarSpectra/@

or:

\verb@      http://hea-www.harvard.edu/QEDT/QuasarSpectra/@.

%%%%%%%%%%%%%%%%%%%%%%%%%%%%%%%%%%%%%%%%%%%%%%%%%%%%%%%%%%%%%%%%%%%%%%%%
%%% figures
%%%%%%%%%%%%%%%%%%%%%%%%%%%%%%%%%%%%%%%%%%%%%%%%%%%%%%%%%%%%%%%%%%%%%%%%

\clearpage

\begin{figure}
% \plotone{f1_1.eps}

Available in:

\verb@      http://lithops.as.arizona.edu/~jill/QuasarSpectra/@

or:

\verb@      http://hea-www.harvard.edu/QEDT/QuasarSpectra/@.

\figcaption{Spectra of quasars obtained with the G130H, G190H or
G270H gratings of FOS.  The 1$\sigma$ error array is plotted as
a dotted line, and the continuum fit is a dashed line.  Absorption
lines more significant than 3.5$\sigma$ are indicated, for
objects whose absorption line spectra have been analyzed.  In some
cases, absorption line analysis was not carried out (see text).  Spectra
have been corrected for Milky Way reddening.}

\end{figure}

\clearpage
\begin{figure}
\plotone{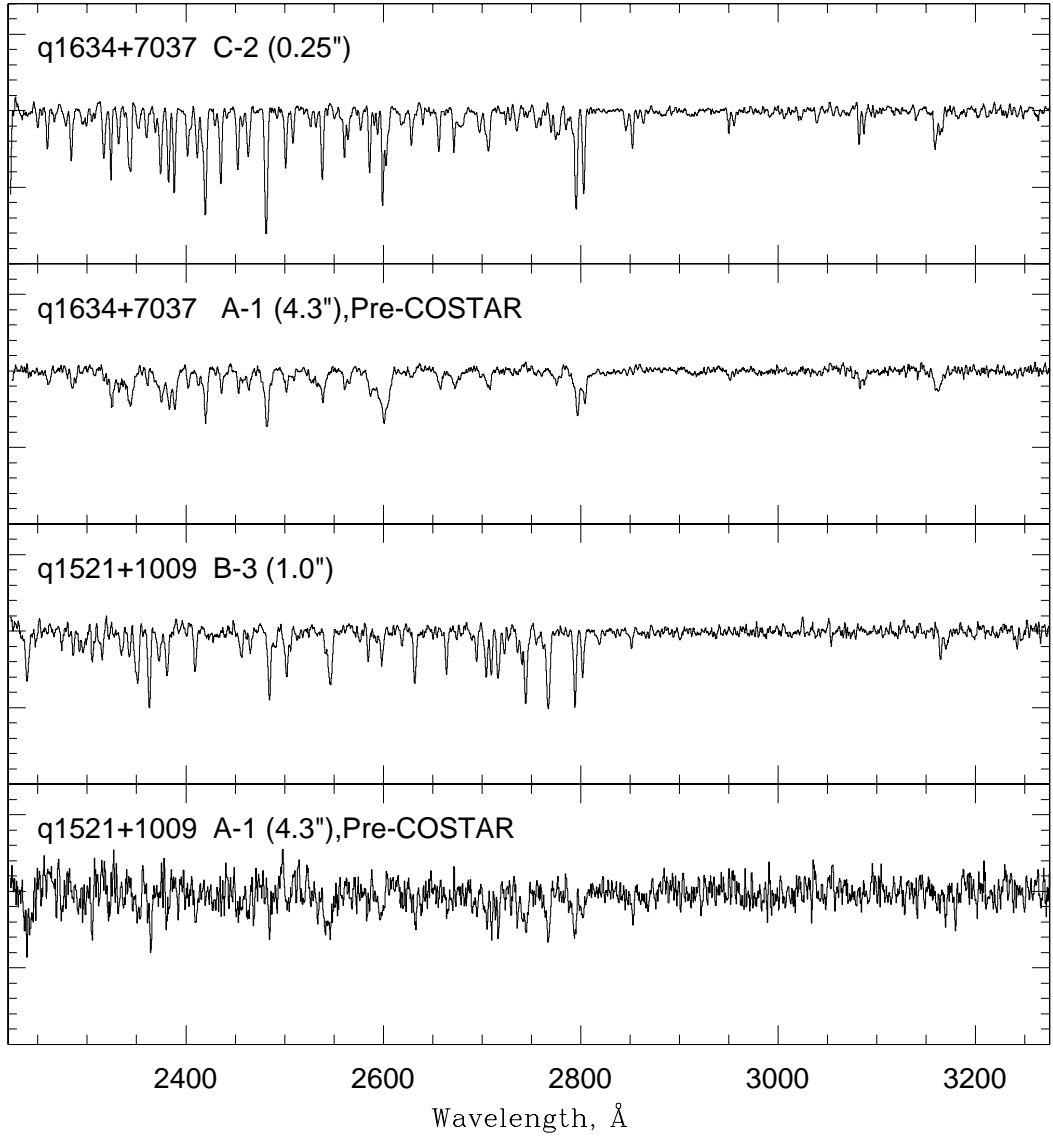}
\figcaption[fig2.ps]{Comparison of A-1 Pre-COSTAR and smaller apertures.  
G270H spectra
of Q 1634+7037 and Q1521+1009 obtained through the A-1 (4.3$\arcsec$)
aperture pre-COSTAR, and the C-2 (0.25 $\arcsec$)  and B-3 (1.0 $\arcsec$) 
apertures, respectively.  The spectra have been normalized to unity by
the continuum fits.
}
\end{figure}

\clearpage
\begin{figure}
\plotone{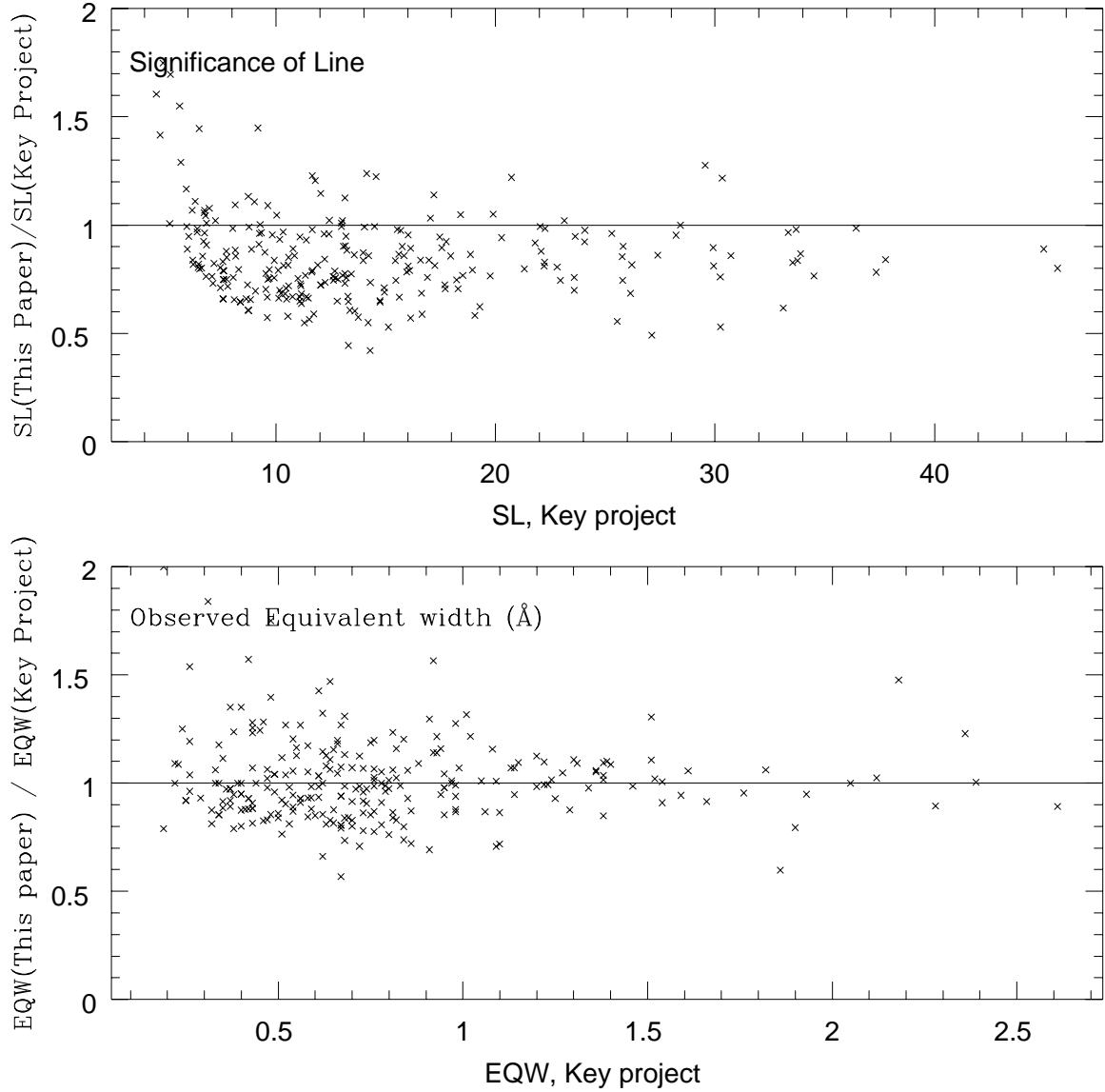}
\figcaption[fig3.ps]{Comparison of our line lists with those of 
the Key Project (J98) for selected objects (see text).  
Top panel: Ratio of the significance assigned to each line by J98 
to that assigned by this work.  
Lower panel: Ratio of the observed equivalent width
of each line measured by J98 to that measured here.
} 
\end{figure}

\end{document}